\documentclass[twocolumn]{aastex63}

\usepackage{graphicx}
\usepackage{subfigure}

\usepackage{hyperref}
\usepackage{natbib}
\usepackage{amsmath}
\usepackage{mathrsfs}

\newcommand{\ageslope}{$m_{\rm logage} = +0.01_{-0.08}^{+0.08}$~}

\newcommand{\fehslope}{$m_{\rm [Fe/H]} = +0.09_{-0.12}^{+0.11}$~}

\newcommand{\mgfslope}{$m_{\rm [Mg/Fe]} = -0.20_{-0.18}^{+0.18}$~}

\begin{document}
\title{Spatially Resolved Stellar Spectroscopy of the Ultra-diffuse Galaxy Dragonfly 44. III. Evidence for an Unexpected Star-Formation History Under Conventional Galaxy Evolution Processes}

\author[0000-0003-1887-0621]{Alexa Villaume}
\altaffiliation{Waterloo Centre for Astrophysics Fellow}
\affiliation{Department of Astronomy \& Astrophysics, University of California Santa Cruz, 1156 High Street, Santa Cruz, CA 95064, USA}
\affiliation{Waterloo Centre for Astrophysics, Department of Physics \& Astronomy, University of Waterloo, 200 University Ave. W., Waterloo, Ontario, Canada N2L 3G1}
\correspondingauthor{Alexa Villaume}
\email{avillaum@uwaterloo.ca}

\author[0000-0003-2473-0369]{Aaron J. Romanowsky}
\affiliation{Department of Physics \& Astronomy, San Jose State University, One Washington Square, San Jose, CA 95192, USA} 
\affiliation{University of California Observatories, 1156 High Street, Santa Cruz, CA, 95064, CA, USA}

\author[0000-0002-9658-8763]{Jean Brodie}
\affiliation{Centre for Astrophysics \& Supercomputing, Swinburne University, Hawthorn VIC 3112, Australia}

\author{Pieter van Dokkum}
\affiliation{Astronomy Department, Yale University, 52 Hillhouse Avenue, New Haven, CT 06511, USA}

\author{Charlie Conroy}
\affiliation{Harvard-Smithsonian Center for Astrophysics, 60 Garden Street, Cambridge, MA, USA}

\author{Duncan A. Forbes}
\affiliation{Centre for Astrophysics \& Supercomputing, Swinburne University, Hawthorn VIC 3122, Australia}

\author{Shany Danieli}
\altaffiliation{NASA Hubble Fellow}
\affiliation{Astronomy Department, Yale University, 52 Hillhouse Avenue, New Haven, CT 06511, USA}
\affiliation{Department of Physics, Yale University, New Haven, CT 06520, USA}
\affiliation{Yale Center for Astronomy and Astrophysics, Yale University, New Haven, CT 06511, USA}
\affiliation{Institute for Advanced Study, 1 Einstein Drive, Princeton, NJ 08540, USA}

\author{Christopher Martin}
\affiliation{Cahill Center for Astrophysics, California Institute of Technology, 1216 East California Boulevard, Mail Code 278-17, Pasadena, CA 91125, USA}

\author{Matt Matuszewski}
\affiliation{Cahill Center for Astrophysics, California Institute of Technology, 1216 East California Boulevard, Mail Code 278-17, Pasadena, CA 91125, USA}

\begin{abstract}
    We use the Keck Cosmic Web Imager integral-field unit spectrograph to: 1) measure the global stellar population parameters for the ultra-diffuse galaxy (UDG) Dragonfly 44 (DF44) to much higher precision than previously possible for any UDG, and 2) for the first time measure spatially-resolved stellar population parameters of a UDG. We find that DF44 falls below the mass--metallicity relation established by canonical dwarf galaxies both in and beyond the Local Group. We measure a flat radial age gradient (\ageslope  log Gyr kpc$^{-1}$) and a flat-to-positive metallicity gradient (\fehslope dex kpc$^{-1}$), which are inconsistent with the gradients measured in similarly pressure-supported dwarf galaxies. We also measure a negative [Mg/Fe] gradient (\mgfslope) dex kpc$^{-1}$ such that the central $1.5$ kpc of DF44 has stellar population parameters comparable to metal-poor globular clusters. Overall, DF44 does not have internal properties similar to other dwarf galaxies and is inconsistent with it having been puffed up through a prolonged, bursty star-formation history, as suggested by some simulations. Rather, the evidence indicates that DF44 experienced an intense epoch of ``inside-out'' star formation and then quenched early and catastrophically, such that star-formation was cut off more quickly than in canonical dwarf galaxies. 
\end{abstract}


\section{Introduction}

While low-surface brightness (LSB) galaxies have been recognized since \citet{sandage1984}, the inherent difficulties in observing them have meant that only a handful of LSB galaxies were studied over decades
\citep[][]{impey1988, bothun1991, dalcanton1997}.
Advancements in telescope design \citep[e.g.,][]{abraham2014} and instrumentation \citep[e.g.,][]{koda2015, yagi2016, vdburg2016} enabled the systematic discovery and analysis of many LSB galaxies.
Of special interest are the ultra-diffuse galaxies (UDGs), with the
luminosities of dwarfs but half-light radii like giants, which led to the initial suggestion that they could be ``failed'' $L^*$, i.e., Milky Way-mass, galaxies \citep{vd2015}.

Subsequent follow-up work focused first on attempting to measure the total masses of UDGs.
This included both dynamical mass measurements, and proxy measurements through the number of globular clusters \citep[GCs, e.g,][]{vd2017b, lim2018}.  
Many UDGs were found to have more populous GC systems than galaxies with similar luminosities, and given empirical scaling relations between GCs and dark matter halos, it was inferred that these UDGs are overmassive for their luminosities  \citep[e.g.,][]{beasley2016a, forbes2020}, albeit not as massive as $L^*$ galaxies.
With implied halo masses of $\sim 10^{11} M_\odot$, these UDGs  could be considered as failed M33-mass galaxies \citep[e.g.,][]{gannon2020}.

UDG formation  in a cosmological context has been simulated  by many groups, with a  diversity  of mechanisms invoked to explain their large sizes, while maintaining standard dark matter halo occupancies
\citep[e.g.,][]{dicintio2017, rong2017, carleton_2020arXiv, tremmel2020, jackson2020, sales2020}.
Thus at first glance it would appear that UDGs are a natural byproduct of conventional processes of galaxy evolution.
However, the models have not yet been able to fully engage with the observed, fundamental properties of UDGs, including not only their sizes and luminosities but also their internal kinematics, stellar populations, and GC systems.
Even the size predictions can be viewed with concern, as models that predict UDGs tend to be unable to reproduce the vast majority of dwarfs with ``compact'' sizes \citep[e.g.,][]{jiang2019}.

As UDGs are studied more systematically, going beyond their basic photometric properties, a complex observational landscape has emerged.
For example, not all UDGs have overpopulous GC systems, with many having more modest populations as expected for dwarf galaxies  \citep[e.g.,][]{somalwar2020, forbes2020}. 
Similarly, the global stellar population parameters of UDGs display a wide range of age, metallicity, and $\alpha$-element enhancement \citep[e.g.,][]{gu2018a, ferre-mateu2018, rl2018, chilingarian2019, mnavarro2019}. 

These observations suggest that ``UDGs'' are an aggregate population of objects with different origins \citep[][]{papastergis2017,forbes2020}.
In this view,  the UDGs that are members of the canonical dwarf population (and predicted by simulations) are those with typical GC populations and stellar populations for dwarfs but still puffed up in some way to have effective radii $\sim 1.5$ kpc.
UDGs that are GC-rich would then comprise the population of  ``failed'' galaxies, with star formation quenched after a very early, rapid formation period, and should therefore also be relatively old, metal-poor, and $\alpha$-enhanced. 
Such a novel and unconventional population of galaxies would not be a mere curiosity, but could provide insights into the earliest stages of galaxy formation, into the origins of GCs, and into the build-up of massive galaxy stellar halos
\citep[e.g.,][]{peng2016, longobardi2018a, villaume2020}.

A prototype for failed UDGs is Dragonfly 44 (DF44), with its high stellar velocity dispersion and populous GC system \citep{vd2016}
\footnote{There are also two well-studied UDGs, NGC~1052-DF2 and NGC~1052-DF4, with extremely {\it low} velocity dispersions that suggest the absence of dark matter \citep{vd2018, vd2019b, danieli2019}. It is not yet clear how these galaxies  fit in with other UDGs, and here we only re-emphasize that any model for their origins must explain their peculiar populations of star clusters, which are  unusually large and luminous for GCs \citep{vd2018b}.}.
It resides toward the outskirts of the Coma cluster 
(clustercentric distance of $\gtrsim 1.8$~Mpc or $\gtrsim 0.6$ virial radii).
DF44 has thus become an important target for further follow-up
to illuminate its nature and origins, including its overall stellar populations, GC system, X-ray emission, and star formation history
\citep{vd2017b,gu2018a,bogdan2020,lee2020,saifollahi2020}.

There has also been a recent push to obtain more detailed information about the internal properties of UDGs, rather than rely only on structural parameters like size, or on proxies like GC systems \citep[e.g.,][]{danieli2019, muller2020}.
Again, DF44 is a key object for analysis, with deep, spatially-resolved spectroscopy made possible by the Keck Cosmic Web Imager \citep[KCWI;][]{morrissey2012, morrissey2018}.
This galaxy's kinematics and dynamics were studied in detail by \citet{vd2019}.
The results on dark matter were unfortunately still not definitive: both standard and overmassive halos were compatible with the data, depending to some extent on the assumed shape of the density profile \citep[see also][]{wasserman2019}.

Other aspects of its kinematics and dynamics highlight DF44's unusual properties, even with respect to other UDGs. 
While \citet{mancera_pina2020} measured high stellar specific angular momentum for a sample of UDGs -- which support the high-spin dark matter halos UDG formation models  \citep{amorisco2016, rong2017} -- \citet{vd2019}  found very low rotation that is in tension with these explanations.
Furthermore, DF44 also appears to be part of a recently accreted, low-mass galaxy group \citep{alabi2018, vd2019} which means that tides from the cluster environment would not have had time to lower a potentially higher original spin.
On the other hand, the radial alignment of UDGs in the Coma cluster \citep[][including DF44]{yagi2016} has been taken as evidence that tidal effects have influenced this cluster UDG population \citep{carleton2019}.

Moreover, there is ongoing debate about the GCs in the DF44 system, with some estimates of their total number being close to standard expectations for dwarf galaxies \citep{saifollahi2020, lee2020}. 
Thus the nature and origins of DF44 remain unsettled, emphasizing the need for fresh observational constraints.

In particular, the internal gradients in the stellar populations of UDGs are essentially uncharted territory.
These gradients probe important processes on sub-kpc scales such as gas accretion and stellar feedback \citep[e.g.,][]{hopkins2018}, 
and so have been used extensively to constrain galaxy formation and evolution scenarios \citep[e.g.,][]{tolstoy2009}.
For UDGs, many of the proposed models involve prolonged and bursty star formation histories \citep[e.g.,][]{dicintio2017, chan2018} that would naturally leave an imprint on the internal gradients.

Here we return to the KCWI data for DF44 obtained by \cite{vd2019}, using detailed stellar population analysis to address outstanding questions about this benchmark UDG.
Is DF44 in some way fundamentally different from the ``canonical'' dwarf population (dwarf ellipticals, irregulars, and spheroidals with more compact sizes)?
And can its evolutionary history be more definitively tested?

In Section~\ref{s.df44.methods} we summarize the salient details of the spectroscopic data and present the results from applying stellar population synthesis models to the data. 
In Section~\ref{s.df44.discussion}, we synthesize our results with other observational constraints to understand the possible formation history of DF44 and contextualize UDGs within the larger framework of galaxy evolution. 
In Section~\ref{s.df44.summary}, we summarize and highlight the key results. 
We assume the distance to the Coma Cluster is 100 Mpc \citep[adopted from][]{liu2001}, corresponding to a distance modulus of 34.99 mag and a scale of 0.474 kpc arcsec$^{-1}$.

\section{Methods and Results}\label{s.df44.methods}

\subsection{Methods}

We make use of the spectroscopic dataset described in \citet{vd2019}, to which interested readers should refer for detailed descriptions of the observations and data reductions. 
Briefly, we obtained integral field unit (IFU) spectroscopy of DF44 with 
KCWI and extracted spectra in nine elliptical apertures following the isophotes of the galaxy. 
Prior to extraction, the 10 brightest point sources in the KCWI field of view were masked.
The remaining point sources comprise only $\sim 2$\% of the total galaxy light.

KCWI enables a huge signal-to-noise (S/N) increase over previous studies of the stellar populations of UDGs. 
For example, \citet{ferre-mateu2018} used DEIMOS, also on Keck II, to obtain integrated spectroscopy for Coma cluster UDGs with S/N that ranges from $\sim 15$--$30$~\AA$^{-1}$ after 14.5 hours of on-target exposure time.
DF44 specifically was included in the Deep Coma project \citep{gu2018a}, 
where 13.5 hours of total on-target IFU integration time with the SDSS telescope
yielded a mean S/N $\sim 7.9$~\AA$^{-1}$ for the integrated spectrum. 
With KCWI, 17 hours of on-target exposure time achieved a S/N $\sim 96$~\AA$^{-1}$ for the integrated spectrum and $12$--$20$~\AA$^{-1}$ for the spatially-resolved spectra.

To extract stellar population parameters from the data, we use the full-spectrum stellar population synthesis (SPS) models \citep[\texttt{alf;}][]{conroy2018a}. 
The most relevant update of \texttt{alf}
is the expansion of stellar parameter coverage of the models with the Spectral Polynomial Interpolator \citep[SPI,][]{villaume2017a}\footnote{\url{https://github.com/AlexaVillaume/SPI_Utils}}. 
With SPI we used the MILES optical stellar library \citep{sb2006a}, the Extended IRTF stellar library \citep[E-IRTF,][]{villaume2017a}, and a large sample of M-dwarf spectra \citep{mann2015} to create a data-driven model from which we can generate stellar spectra as a function of effective temperature ($T_{\rm eff}$), surface gravity, and metallicity.

The empirical parameter space is set by the E-IRTF and \citet{mann2015} samples which together span $-2.0 \lesssim {\rm [Fe/H]} \lesssim +0.5$ and $3.9 \lesssim {\rm log}~T_{\rm eff} \lesssim 3.5$. 
To preserve the quality of interpolation at the edges of the empirical parameter space, we augment the training set with a theoretical stellar library (C3K). 
The \texttt{alf} models allow for variable abundance patterns by differentially including theoretical element response functions. 
We use the measured Mg abundances for the MILES stellar library stars from \citet{milone2011} to derive the [Mg/Fe] versus [Fe/H] relation in our model. 

To fit the models to data we employ a Markov chain Monte Carlo algorithm \citep{emcee_v1} to sample the posteriors for the stellar population parameters (see Table 2 in \citealt{conroy2018a}). 
The ``best-fit'' value for each parameter is taken as the median (50th percentile) of the corresponding posterior, while the uncertainty is the 1$\sigma$ credible interval, i.e., the 18th and 84th percentiles.

We used the Medium slicer on KCWI, yielding a spectral resolution of $R\sim4000$. 
This necessitated smoothing the data to the native resolution of the \texttt{alf} models, $110$ km~s$^{-1}$.
It has been previously demonstrated that the accuracy of the recovered stellar parameters is not affected by smoothing \citep[see Appendix A in][]{choi2014}.
However, the possibility of pixels affected by sky lines being smeared throughout the spectra from smoothing is a particular concern for low-surface brightness observations. 
To mitigate this effect, we interpolated over the pixels badly affected by sky transmission before smoothing to the desired velocity resolution by convolving a wavelength dependent Gaussian kernel with $\sigma = \sqrt{\sigma_D^2 - \sigma_I^2}$, where $\sigma_D$ is the desired resolution and $\sigma_I$ is the instrumental resolution \citep[see Figure 9 in][]{vd2019}. 

\begin{figure*}
    \includegraphics[width=1\textwidth]{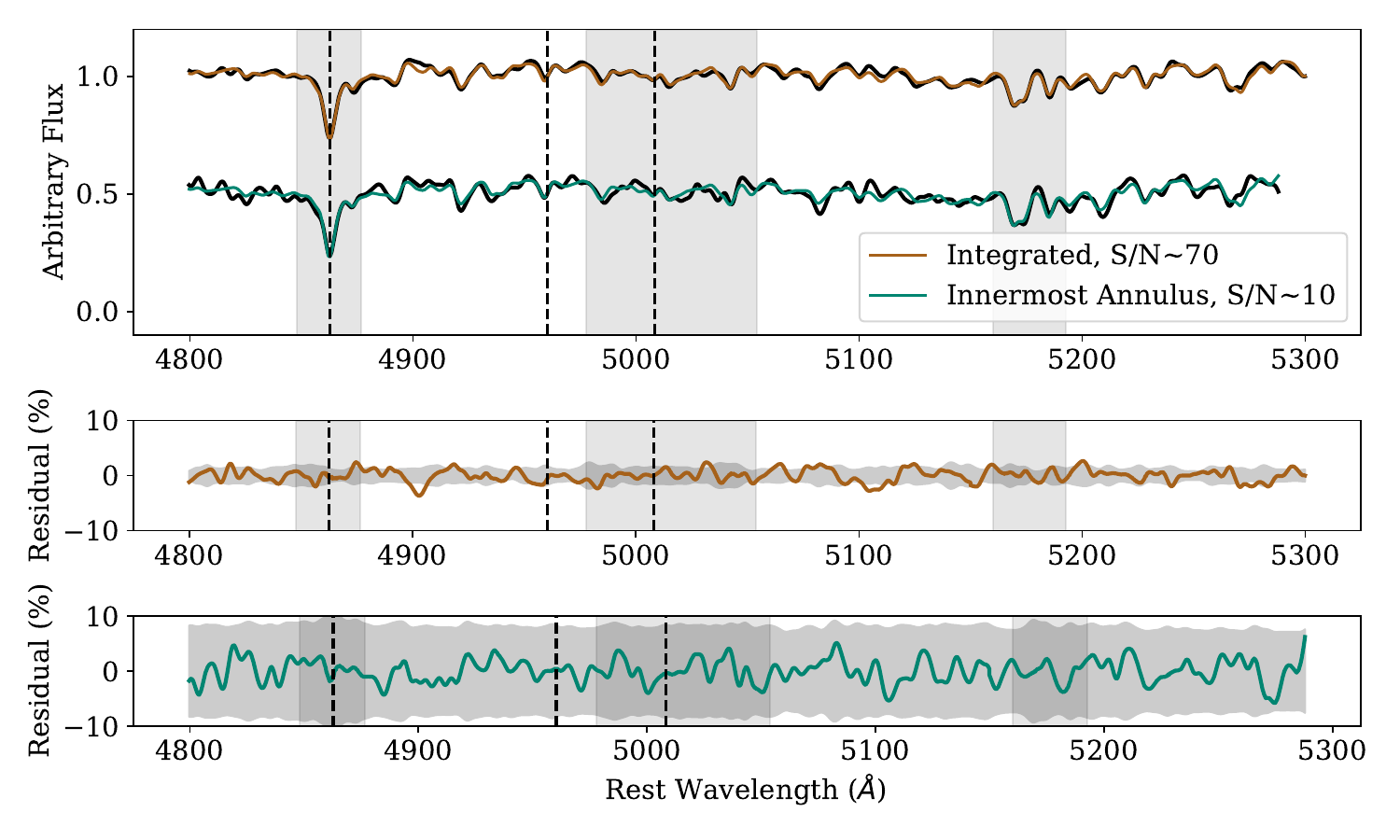}
    \caption{Top: comparison of observed DF44 KCWI spectra (black) and best-fit models for the integrated spectrum (brown) and the annulus with the lowest S/N (green). Middle: comparison of the residuals between the best-fit model and data for the integrated spectrum, and uncertainty on the flux (gray). 
    Bottom: same as middle panel but for the spatially-resolved spectrum.  
    The locations of H$\beta$, Fe5017, and Mg$b$ are highlighted in the middle and bottom panels (vertical gray regions) as well the locations of possible emission lines (dashed lines) but, as explained in Section 2.2.1, there is no evidence of emission lines in the spectra.}
    \label{fig:model_check}
\end{figure*}

We checked for a bias potentially introduced by the smoothing by fitting the spectra smoothed to resolutions ranging from $100 \leq \sigma \leq 200$ km~s$^{-1}$. 
The $\sigma$-dependent differences in the inferred stellar population parameters (not shown) were much smaller than the uncertainties on those parameters. 
We also tried several interpolation techniques to remove the bad pixels and found little sensitivity in the final parameters.   

We fit over the wavelength regions $4800 \leq \lambda {\rm \AA (obs)} \leq 5150$ and $5150 \leq \lambda {\rm \AA (obs)}\leq 5300$. 
We use the standard priors \citep[see][for details]{conroy2018a}, except for the prior on [Mg/Fe]. Based on examination of the posteriors, we changed the lower prior to be [Mg/H]$=-1.0$ for the final fits presented in this work.

We examine the agreement between the observed spectrum (black) and the best-fit (brown) model for the global  spectrum in the top panel of Figure \ref{fig:model_check}.
In the middle panel we compare the agreement between the residuals between the data and best-fit model for the global spectrum with the uncertainty on the data (gray band). 
In the top and bottom panels of Figure \ref{fig:model_check} we make the same comparison for the spatially-resolved spectrum with the lowest S/N ($\sim 10$, green). 
Also marked are key spectral features. 
Within the uncertainties, the best-fit models reproduce the key characteristics of both the integrated and spatially-resolved spectra.

\subsection{Global Properties}

\begin{figure}
    \centering
    \includegraphics[width=0.5\textwidth]{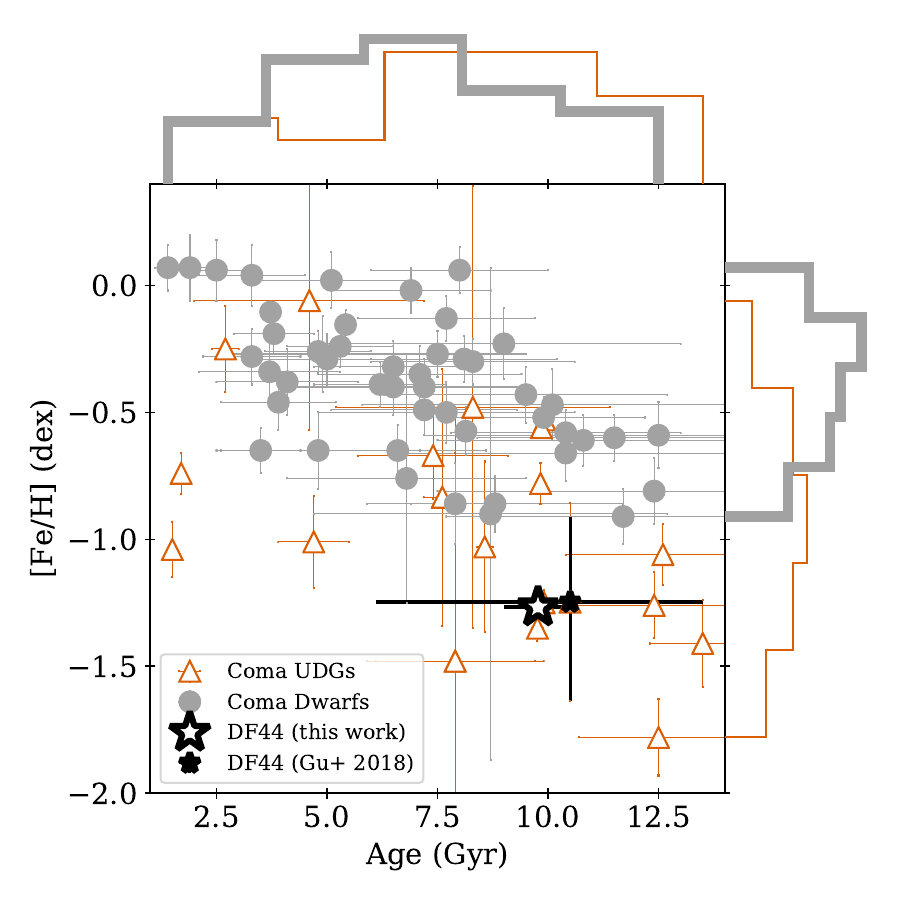}
    \caption{Comparing the relation between global age and [Fe/H] for a compilation of Coma UDGs (orange) and typical Coma dwarf galaxies (grey). The measurements for DF44 are highlighted from both this work (large open star) and from \citet{gu2018a} (filled black star). Also shown are the marginal distributions of age and [Fe/H] for both the UDG and dwarf samples. There is small difference in the age distributions of the two populations but the UDG sample is overall more metal-poor than the sample of canonical dwarf galaxies. 
    }
    \label{fig:gp_age}
\end{figure}

In Figure \ref{fig:gp_age} we show the relation between age and [Fe/H] for a sample of canonical Coma dwarf galaxies \citep[grey circles;][]{smith2009, gu2018a, ferre-mateu2018}  and a sample of Coma UDGs \citep[orange triangles;][]{ferre-mateu2018, rl2018, gu2018a, chilingarian2019}\footnote{We defer to the galaxy classifications of the original papers. It should be noted, however, that consistent application of the canonical definition of UDGs \citep[semi-major axis effective radius $\gtrsim 1.5$ kpc,  mean surface brightness $\langle \mu_g \rangle_{\rm e} > 25$ mag arcsec$^{-2}$;][]{vd2015} is not used in all these works. 
There is also some overlap in the objects included in these papers. 
Neither of these things has any bearing on the analysis presented in this work.
}. 
The result of modeling the integrated light spectrum of DF44 in this work is also shown (open star). 
We highlight the measurements made for DF44 by \citet{gu2018a} (filled star) in this figure.
The \citet{gu2018a} measurements were made from completely independent data but were also measured using \texttt{alf}.

\begin{table}[]
\begin{tabular}{llll}
\hline
Percentile & log Age (Gyr) & [Fe/H] & [Mg/Fe] \\
 \hline
 \hline
16 & $0.97$ & $-1.37$ & $-0.16$ \\
50 & $1.01$   & $-1.33$ & $-0.10$ \\
84 & $1.04$ & $-1.28$ & $-0.04$ \\
\hline
\end{tabular}
\caption{\label{table:global} The 16th, 50th, and 84th percentiles of the global stellar population parameters measured from the integrated KCWI spectrum of DF44.}\label{table:alf_results}
\end{table}

We also show the marginal age and [Fe/H] distributions for both the dwarf and UDG samples on the x and y axes, respectively. 
While there is no significant difference in the age distributions between the dwarf and UDG samples, as a population the UDGs are more metal-poor than the dwarfs, with some overlap in the two samples. 
DF44 is among the most metal-poor UDGs, more metal-poor than any of the canonical dwarf galaxies, and comparatively old.

\begin{figure}
    \centering
    \includegraphics[width=0.5\textwidth]{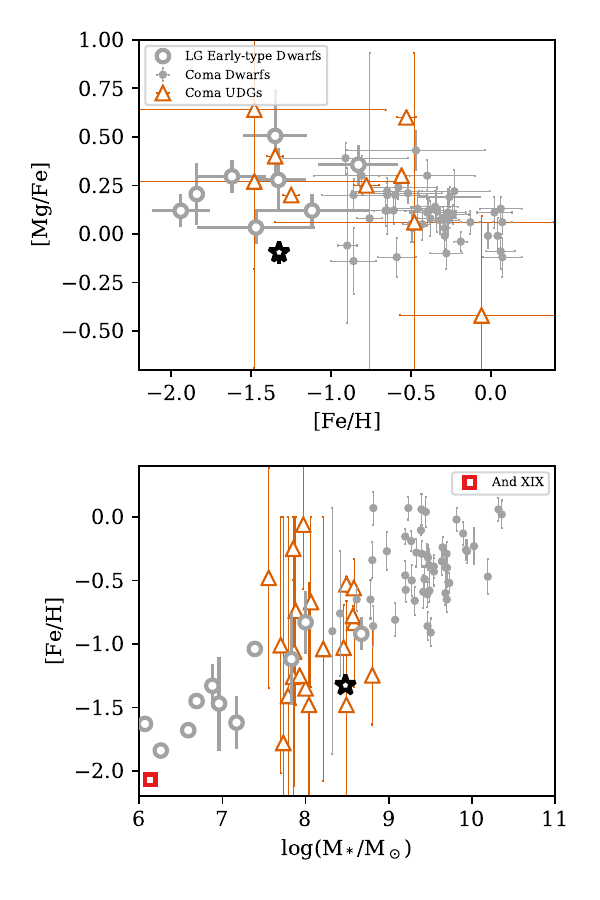}
    \caption{Top: comparing the relation between global [Fe/H] and [Mg/Fe] of Coma UDGs and dwarfs (symbols same as Figure~\ref{fig:gp_age}) as well as Local Group early-type galaxies (open grey circles). Bottom: the mass--metallicity relation for the same objects as the top panel. The Local Group dwarf which has been suggested as a nearby analogue to UDGs, And XIX, is highlighted (red).
    }
    \label{fig:gp_feh}
\end{figure}

In the top panel of Figure \ref{fig:gp_feh}, we compare the relationship between [Fe/H] and [Mg/Fe] for the Coma dwarfs and UDGs, and we now also include a sample of early-type dwarf galaxies in the Local Group \citep{kirby2013, vargas2014}. 
The Coma dwarfs become less Mg-enhanced with increasing [Fe/H], but the Local Group dwarfs do not show such a trend.
The UDGs span the full range of Mg and Fe enhancements of the combined Local Group and Coma cluster sample, but with the paucity of UDGs with measured Mg abundances and large uncertainties, it is presently impossible to discern whether or not there is a UDG-specific trend.
DF44 has a sub-solar Mg-abundance, unlike the other objects with similarly low-metallicity.

However, we urge caution when drawing conclusions from a set of heterogeneously measured $\alpha$ abundances.
\citet{rong2020} noted a $\sim 0.3$ dex offset in their measurements of [Mg/Fe] for NGC 1052-DF2 using the \texttt{MILES} SSPs \citep{vazdekis2015} and the \citet{thomas2011} SSPs.
Measuring $\alpha$ abundances from SPS models has been a historically difficult problem, primarily (but not exclusively) due to the [$\alpha$/Fe] bias introduced in the models from the empirical stellar libraries \citep[see][for a review]{conroy2013}.
However, the most recent SPS models have been converging on similar solutions to this problem (we refer interested readers to the original SPS papers for details) and we note that significant progress has been made in bringing different SPS models in agreement with one another \citep[Figure 16;][]{conroy2018a}.
It is therefore difficult to understand the large offset reported by \cite{rong2020}.
Assessing this is beyond the scope this paper and so we refrain from drawing any conclusion based on comparing our measurement of [Mg/Fe] for DF44 to the published literature values (\citeauthor{gu2018a} sample is not included on this panel as they did not measure [Mg/Fe]).

In the bottom panel of Figure \ref{fig:gp_feh}, we show the mass--metallicity relation for these same samples, and now we include Andromeda XIX (open red circle), the Local Group galaxy suggested by \citet{collins2020} as a local UDG analogue.
The sources for the stellar masses, or the methods for estimating them, are described in Appendix \ref{ap:masses}.

There is some overlap in stellar mass between the Local Group and Coma dwarfs and the two samples appear to follow a continuous mass--metallicity relation. The UDG, sample, however, shows significant scatter in [Fe/H] over a narrow mass range.
DF44 has low metallicity for its mass compared to Local Group and Coma dwarfs.
To quantify DF44's deviation from the mass--metallicity relation we measured the mean metallicity ($\mu_{\rm [Fe/H]} = -0.61$) and standard deviation ($\sigma_{\rm [Fe/H]} = 0.31$) of the 10 dwarfs with $8 \leq {\rm log~M_* (M_\odot)} \leq 9$.
With a stellar metallicity of [Fe/H] $ = -1.33$ this places DF44 at $\sim 2.3\sigma$ below the canonical dwarf relation.

We emphasize that the causes for concern in comparing heterogeneously measured [Mg/Fe] values are not relevant when comparing [Fe/H].
However, an important consideration while comparing our measurements of DF44 with the literature values are the different apertures over which the measurements are taken.
The Coma measurements are from longslit observations, while the DF44 and Local Group measurements integrate spatially-resolved values into a single, global value. 
However, as we show in Section \ref{sec:gradients}, the nature of the stellar metallicity gradient for DF44 demonstrates that confining our ``global'' [Fe/H] measurement to only its central region (i.e., to make it more directly comparable to the Coma measurements) would not help mitigate its discrepancy from the mass--metallicity relation and, in fact, would exacerbate it.

The results for the global stellar population parameters for DF44 are summarized in Table~\ref{table:global}.

\subsubsection{Lack of Evidence for Recent Star Formation}

The results presented thus far are from using the ``simple'' model in \texttt{alf}.
This fixes the stellar initial mass function (IMF) at \citet{kroupa2001} and fits for light-weighted ages and abundances. 
Since age can have a non-linear effect on an integrated spectrum \citep{serra2007} the mass-weighted age could be substantially different than the light-weighted age, if there is a significant young population due to an extended star formation history (SFH).

\begin{figure}
    \includegraphics[width=0.5\textwidth]{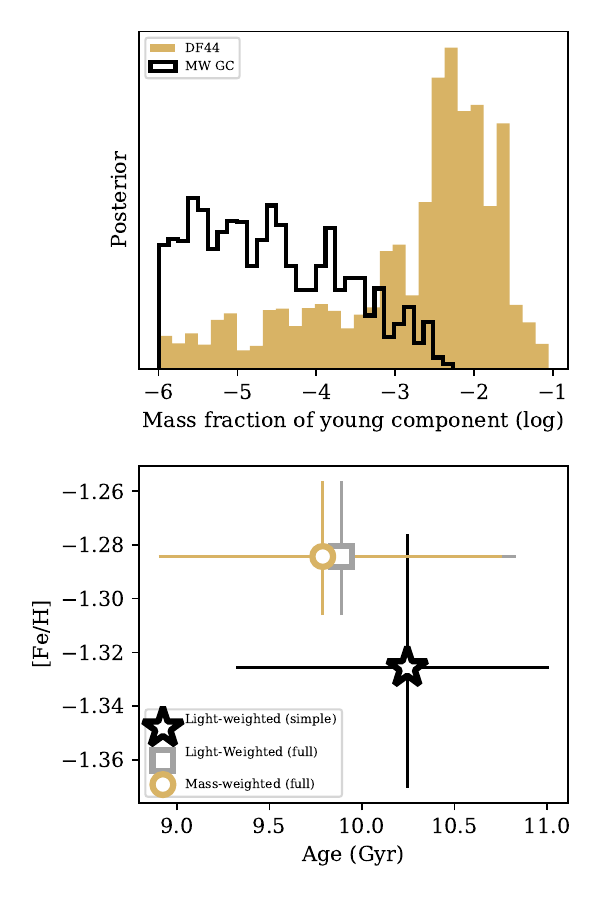}
    \caption{Top: Comparing the inferred mass fractions of a young stellar component for the optimally-extracted DF44 spectrum and a Milky Way GC when fitted with a two-component SFH. 
    While it appears that DF44 may not have an entirely coeval stellar population, any young ($1-3$ Gyr) population has a negligible impact on the integrated spectrum.
    Bottom: Going from the simple (black star) to the full (gray square) \texttt{alf} model produces a negligible impact on recovered light-weighted age and stellar metallicity.
    Moreover, the mass-weighted age (yellow circle) is consistent with the light-weighted age.}
    An ongoing bursty SFH is a proposed mechanism to create UDGs, but in DF44 we find no evidence for recent star formation.
    \label{fig:sfh}
\end{figure}

The integrated spectrum has sufficiently high S/N to use the ``full'' model in \texttt{alf}.
Among other things, the full model fits for a two-component SFH.
This works by fitting for a separate mass fraction and age of a young ($1-3$ Gyr) stellar component.
We show the posterior of the mass fraction of the young component for DF44 (yellow) in the top panel of  Figure~\ref{fig:sfh}.
To contextualize this result we also show the same for a Milky Way GC fitted in the same manner as DF44 \citep[NGC 6638, black; spectrum from][]{schiavon2005}.
The Milky Way GCs have measured old ages \citep[$\gtrsim 10$ Gyr;][]{vandenberg2013} and are the closest known objects to simple stellar populations (that is, formed from a single burst of star formation). 
While DF44 has a higher young mass fraction overall than the GC (as expected from an object that is not a simple stellar population), the mass fraction of the young component for DF44 is still negligible.

In the bottom panel of Figure \ref{fig:sfh} we compare the light-weighted age from the simple model (i.e., same as in Figure \ref{fig:gp_age}; black diamond) with the light-weighted and mass-weighted ages from the full model (gray square and yellow circle respectively).
The ages (and metallicities) are all consistent with one another.

Moreover, when fitting with the full \texttt{alf} model, we also fit for the strength of several emission lines. 
There is no evidence of emission lines in the spectra (location of emission  lines are marked with dashed lines in Figure~\ref{fig:model_check}) and all inferred emission line strengths were insignificant (not shown).

\subsection{Spatially-resolved Stellar Population Measurements}\label{sec:gradients}

\begin{figure*}[ht]
		\includegraphics[width=1.0\textwidth]{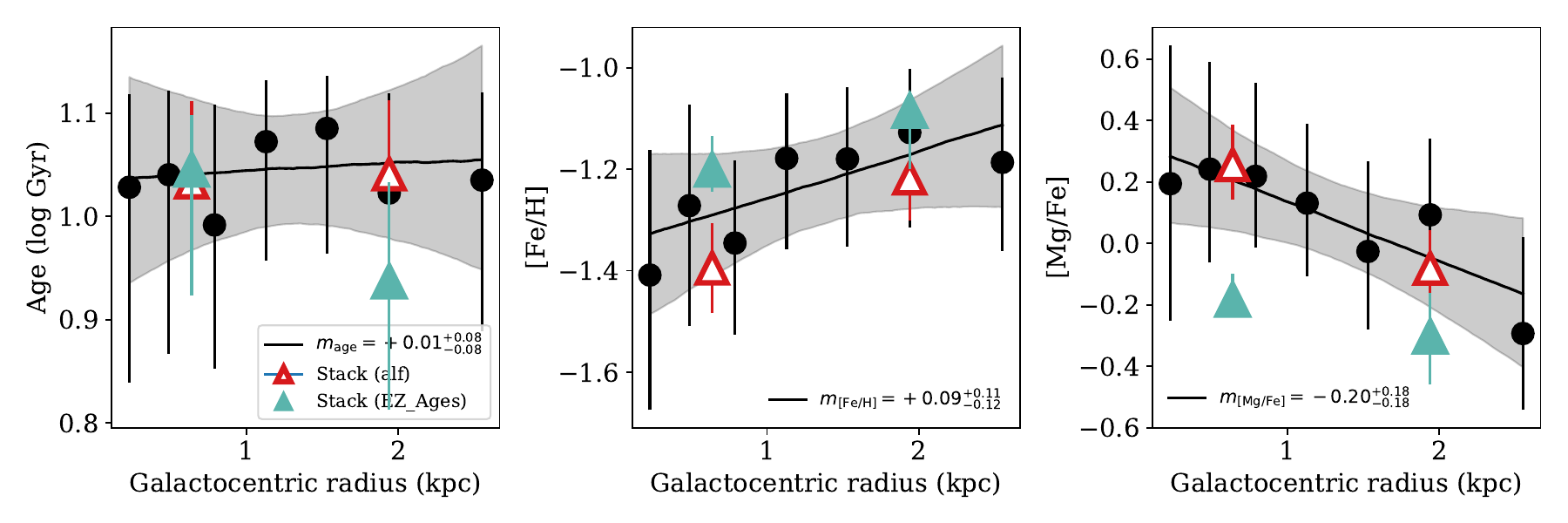}
		\caption{\label{stellar_pop_gradient} Radial profiles of DF44 stellar population parameters (black circles). Also shown are the measurements of two radial bins using both \texttt{alf} (open red triangles) and \texttt{EZ\_Ages} (filled blue triangles). From left to right: log stellar age, [Fe/H], and [Mg/Fe].
		These results suggest that DF44 formed ``inside-out'' (see Section 3.2 for details).
		}
\end{figure*}

In Figure~\ref{stellar_pop_gradient} we show the spatially-resolved measurements as a function of galactocentric radius for each parameter (black circles), from left to right: stellar age, [Fe/H], and [Mg/Fe].

\begin{table}[]
\begin{tabular}{ccccc}
\hline
\hline
Aperture & $R$ & log Age  & [Fe/H]  & [Mg/Fe] \\
         & (kpc) & (Gyr)  & (dex)  & (dex) \\
\hline
$0$ & $0.23$ & $0.84$ & $-1.67$ & $-0.25$ \\
$1$ & $0.49$ & $0.87$ & $-1.51$ & $-0.06$ \\
$2$ & $0.79$ & $0.85$ & $-1.53$ & $-0.01$ \\
$3$ & $1.13$ & $0.96$ & $-1.36$ & $-0.11$ \\
$4$ & $1.53$ & $0.96$ & $-1.35$ & $-0.28$ \\
$5$ & $1.94$ & $0.87$ & $-1.31$ & $-0.17$ \\
$6$ & $2.55$ & $0.89$ & $-1.36$ & $-0.54$ \\
\hline
\end{tabular}
\caption{\label{table:resolved_16} The 16th percentiles of the stellar population parameters measured from the spatially-resolved spectra.}
\end{table}

\begin{table}[]
\begin{tabular}{ccccc}
\hline
Aperture & $R$ & log Age  & [Fe/H]  & [Mg/Fe] \\
         & (kpc) & (Gyr)  & (dex)  & (dex) \\
\hline
\hline
$0$ & $0.23$ & $1.03$ & $-1.41$ & $0.19$ \\
$1$ & $0.49$ & $1.04$ & $-1.27$ & $0.24$ \\
$2$ & $0.79$ & $0.99$ & $-1.35$ & $0.22$ \\
$3$ & $1.13$ & $1.07$ & $-1.18$ & $0.13$ \\
$4$ & $1.53$ & $1.09$ & $-1.18$ & $-0.03$ \\
$5$ & $1.94$ & $1.02$ & $-1.13$ & $0.09$ \\
$6$ & $2.55$ & $1.03$ & $-1.19$ & $-0.29$ \\
\hline
\end{tabular}
\caption{\label{table:resolved_50} The 50th percentiles of the stellar population parameters measured from the spatially-resolved spectra.}
\end{table}

\begin{table}[]
\begin{tabular}{ccccc}
\hline
\hline
Aperture & $R$ & log Age  & [Fe/H]  & [Mg/Fe] \\
         & (kpc) & (Gyr)  & (dex)  & (dex) \\
\hline
$0$ & $0.23$ & $1.12$ & $-1.16$ & $0.64$ \\
$1$ & $0.49$ & $1.12$ & $-1.07$ & $0.59$ \\
$2$ & $0.79$ & $1.11$ & $-1.18$ & $0.52$ \\
$3$ & $1.13$ & $1.13$ & $-1.05$ & $0.39$ \\
$4$ & $1.53$ & $1.14$ & $-1.04$ & $0.27$ \\
$5$ & $1.94$ & $1.12$ & $-1.00$ & $0.34$ \\
$6$ & $2.55$ & $1.12$ & $-1.02$ & $0.02$ \\
\hline
\end{tabular}
\caption{\label{table:resolved_84} The 84th percentiles of the stellar population parameters measured from the spatially-resolved spectra.}
\end{table}

We used a linear regression model accounting for the uncertainties on the individual data points to measure the slopes of each gradient: 
\ageslope log dex kpc$^{-1}$, \fehslope dex kpc$^{-1}$, and \mgfslope dex kpc$^{-1}$. We show the ``best-fit'' lines in Figure~\ref{stellar_pop_gradient} (black lines) and the ranges between the 16th and 84th percentiles (grey bands). 

The stellar population gradients we measure for DF44 are not typical of early-type dwarf galaxies (see further discussion on this in Section \ref{s.df44.discussion}). 
To check our results we created two spatial bins, where the first includes the inner four apertures and the second includes the outer three apertures, achieving S/N $\sim 30~{\rm \AA^{-1}}$ for each stack. 
The stacks were created by bootstrapping the individual spectra included in each bin. 
We used the 50th percentile of the resulting flux distribution as the stacked spectrum, and the average of the 16th and 84th percentiles as the uncertainty spectrum.

We fit these spatial stacks using \texttt{alf} (open red triangles in Figure~\ref{stellar_pop_gradient}) and find that the measurements from the spatial stacks are consistent with both the overall values and the directions of the gradients.

Additionally, we show the result of fitting the spatial stacks with \texttt{EZ\_Ages} (filled turquoise triangles])\footnote{We measured the H$\beta$, Fe5015, Mg$b$, C$4668$, CN$2$, and Ca$4227$ indices after correcting the spectra to rest-frame using the \texttt{alf} measurements of recession velocity for each. For the fitting, we used the solar-scaled isochrones, an IMF exponent of $1.35$, and [Ti/Fe] $=0$.} to check for possible model-based systematics, as \texttt{EZ\_Ages} is an index-based SPS model using an entirely different stellar library and isochrone grid. 
We fit the full flux distribution for each stack from the bootstrapping described above. 
The best-fit values and uncertainties are the median and 1$\sigma$ uncertainties from those fits.

We also fit the integrated spectrum with \texttt{EZ\_Ages} (not shown) and found good agreement for stellar age and [Fe/H] with the \texttt{alf} measurements. 
For the spatial stacks,
the best-fit ages are consistent between \texttt{EZ\_Ages} and \texttt{alf}.  \texttt{EZ\_Ages} is based on a more limited stellar spectral library than \texttt{alf} and is not reliable for abundance measurements where [Fe/H] $\lesssim -1.0$ \citep[see][]{schiavon2013}. 
While there is broad agreement in the [Fe/H] measurements, the [Fe/H] measurements for the inner spatial stack are not consistent with each other within the uncertainties. This may also explain the discrepancy in the [Mg/Fe] measurements. 
In any case, the direction of the stellar population gradients as measured from \texttt{EZ\_Ages} is consistent with the \texttt{alf} measurements, indicating that model-based systematics are not erroneously introducing the radial trends.

\begin{figure}
    \includegraphics[width=0.5\textwidth]{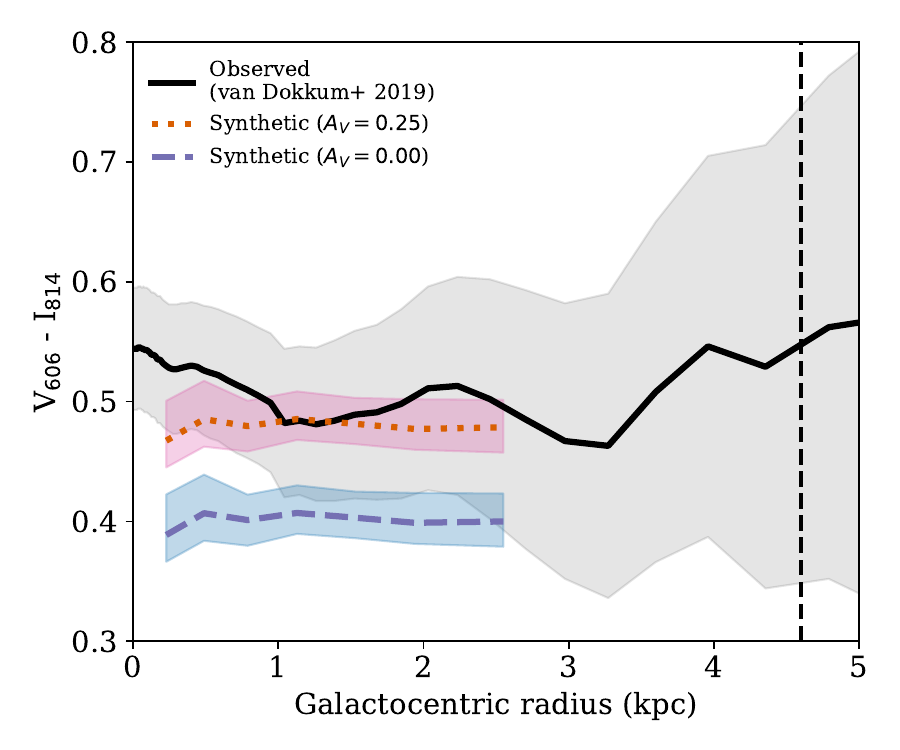}
    \caption{Comparison of observed DF44 color profile (gray) to synthetic color profiles generated from the best-fit models of the spatially-resolved spectra without dust extinction (blue) and with dust extinction (red). The overall shape of the synthetic color profile is consistent with the observed color profile but there is a small offset in overall color. Independent analysis of near-infrared photometry indicates there may be a dust component in DF44. The color profile does not change appreciably between the outer range of the spectroscopic coverage and the half-light radius (black vertical dashed line). }
    \label{color_profile}
\end{figure}

As an additional check on the veracity of our gradient measurements, we computed a $V_{606}-I_{814}$ color profile from synthetic spectra generated from the distribution of stellar parameters inferred from the \texttt{alf} models to compare to the observed color profile\footnote{The observed color profile presented here is derived from an ellipse fit on the smoothed {\it HST} images. 
It is an independent analysis from what was published in \citet{vd2019}, 
with results that are essentially identical except for larger uncertainties (these are dominated by the sky subtraction, and are correlated in radius).} (blue and black lines, respectively, in Figure~\ref{color_profile}). 
We also mark the effective radius of DF44, $R_{\rm e} = 4.6 \pm 0.2$ kpc, in this figure (dashed vertical black line).
The overall behavior of the synthetic profile is consistent within the large uncertainties of the observed profile but there is a small overall offset ($\sim 0.1$ mag) between the synthetic and observed color profiles. 
This is possibly due to issues in the isochrones for the upper giant branch \citep{choi2016}, although synthetic colors have been generated for globular clusters in the Milky Way and M87 \citep[see][]{villaume2019} without this discrepancy. 

Another possibility is the presence of dust in DF44. 
Independent work modeling DF44 broadband photometry from the optical to the near-infrared ($0.3 - 5 {\rm \mu m}$) inferred an overall dust attenuation of $A_V \sim 0.25$ (S.~Laine, private communication). 
We show that applying this dust attenuation value to the synthetic color profile (red) resolves the discrepancy in Figure~\ref{color_profile}.
This raises the question of the origin of this potential dust component, which we defer to future work.

\begin{figure}
    \includegraphics[width=0.5\textwidth]{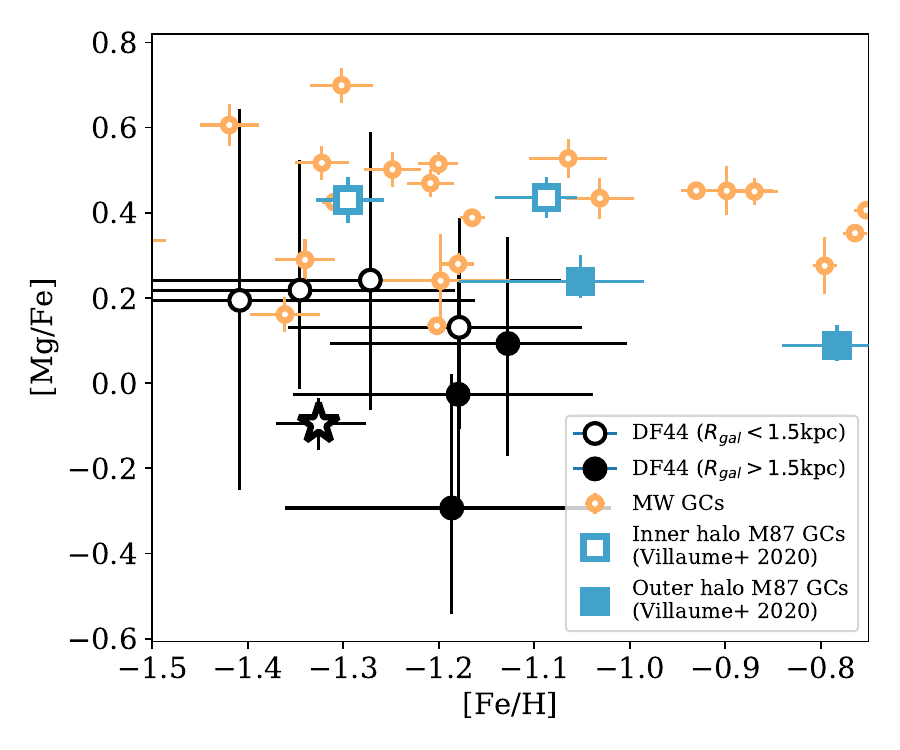}
    \caption{Comparing the spatially-resolved measurements of  DF44 to spatially-resolved measurements from spectral stacks of M87's inner halo GC population ($R_{\rm gal} < 40$ kpc, open blue squares) and outer halo GC population ($ 40 < R_{\rm gal}\text{ kpc} < 140$), filled blue squares), and individual Milky Way GCs (open orange circles). The DF44 measurements are split between the inner sample ($R_{\rm gal} < 1.5$ kpc, open black circles) and the outer sample (filled black circles). The similarity between DF44's central region and the metal-poor GC subpopulation suggest a potentially coeval evolution.}
    \label{fig:chemo_evo}
\end{figure}
 
In Figure \ref{fig:chemo_evo} we look at [Fe/H] vs. [Mg/Fe] measured from the spatially-resolved DF44 spectra. We indicate the apertures that are within $1.5$ kpc as open black circles, and those farther out as filled black circles. We compare the measurements for DF44 to a sample of individual Milky Way GCs \citep[fitted from integrated spectra obtained by][open orange circles]{schiavon2005} and spatially-resolved stacks of M87 GCs \citep[][blue squares]{villaume2020} demonstrating that the stellar populations of the inner core of DF44 have similar properties to metal-poor GCs.

\section{Discussion}\label{s.df44.discussion}

\subsection{How Similar is the Evolutionary History of DF44 to Canonical Dwarfs?}

\subsubsection{Comparison to observations}
\label{sec:comparison_to_observations}
To help clarify the origins of DF44, we first examine whether or not  its stellar population gradients are similar to those of  galaxies of the same luminosity but more average (compact) sizes.
One complication in attempting to answer this question is the significant diversity in the properties of dwarfs -- from gas rich star-forming dwarf irregulars (dIrrs) to gas poor quiescent dwarf spheroidals (dSphs) and dwarf ellipticals \citep[dEs; see the review by][]{tolstoy2009}. 
The dEs further display a range in kinematic properties with some that are rotation-supported, and possibly the quenched descendants of dIrrs, and there are those that are pressure-supported, which do not have an established formation pathway \citep{vanzee2004}.

\begin{figure}[ht]
		\includegraphics[width=0.5\textwidth]{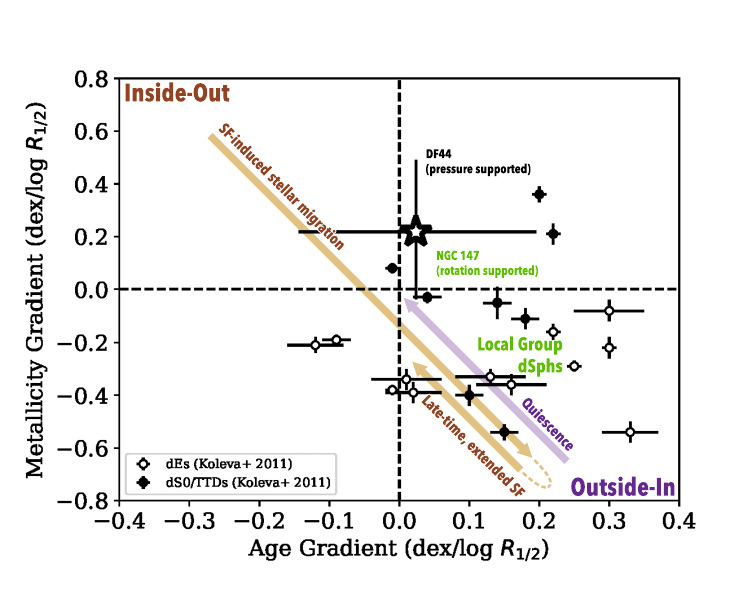}
		\caption{\label{fig:gradient_context} Comparing the age and metallicity gradients we measure for DF44 with the sample of dEs (open circles) and dS0s/TTDs from \citet{koleva2011} (filled circles). 
		In green text are objects which can only be qualitatively shown on this figure.
		The yellow (``inside-out'') and purple (``outside-in'') arrows show two general models for
		how age and metallicity gradients arise in dwarf galaxies. 
		Both processes have been invoked to explain the gradients of the Local Group dSph population (i.e., positive age gradients and negative metallicity gradients: see text in Section \ref{s.df44.discussion} for more discussion and detailed references).
		Whether star formation proceeds in an inside-out or outside-in manner, the old age and the nearly flat gradients we measure for DF44 are not consistent with it having an extended star formation history, disfavoring the formation scenario that invokes ongoing stellar feedback to explain its large size.}
\end{figure}

DF44 now has the deepest spectroscopy, and consequently the best analysis of its dynamics and stellar populations, of any UDG.
In Figure \ref{fig:gradient_context} we summarize current observations of age and metallicity gradients in dwarf galaxies, and how they relate to what we measure for DF44. 
For a quantitative comparison we use the sample of gradients presented in \citet{koleva2011}.
We separate the dEs (open circles) from this sample as they are the most morphologically and dynamically similar to DF44 (i.e., pressure supported). 
We also show the transition type dwarfs (TTDs) and the dwarf lenticulars (dS0s), i.e., the dwarfs that show evidence for disk features or being rotation supported (filled circles).

The age and metallicity gradients of the dEs in \citet{koleva2011} are similar to the Local Group dSphs. 
In this figure we can only indicate the Local Group galaxies qualitatively, but the age and metallicity gradients seem to be near-universal, with extended populations of old, metal-poor stars surrounding centrally concentrated populations of young, metal-rich stars evident in imaging surveys \citep[e.g.,][i.e., the lower right quadrant of Figure \ref{fig:gradient_context}]{harbeck2001, tolstoy2004, faria2007, kacharov2017}.
The metallicity and $\alpha$ gradients measured from spectroscopy in the Local Group, while still comparatively rare, are consistent with the picture indicated by the imaging \citep[e.g.,][]{kirby2009, leaman2013, vargas2014}.

We note that NGC 147, a Local Group dE, is an exception, with a flat to slightly positive metallicity gradient \citep[][]{vargas2014}. 
It is rotation-supported \citep{geha2010}, and fits the pattern from the
\citet{koleva2011} sample of only (some) rotation-supported dwarfs having positive gradients.
The Local Group dwarf irregular WLM, which may be a star-forming UDG analog (e.g., \citealt{bellazzini2017}), has a mildly negative metallicity gradient, and a positive age gradient \citep{leaman2013,albers2019}.


The age and metallicity gradients we measure for DF44 are unusual by comparison.
While DF44 is morphologically and dynamically similar to dSphs and dEs, it does not share the distinct stellar population gradients of those galaxies.
And while the metallicity gradient of DF44 is similar to those measured in some dS0s and TTDs, it does not share the same characteristic rotation-supported kinematics.
Furthermore, the Local Group dwarfs typically have flat-to-positive $\alpha$ gradients \citep{vargas2014,hayes2020}, while we see a decline in [Mg/Fe] with increasing radius in DF44.

\subsubsection{Comparison to formation scenarios}

Many scenarios have arisen from simulations and semi-analytic models that show that it is possible to mimic some of the basic characteristics of the UDG population without significant adaptations to the formation and evolutionary physics thought to create the canonical dwarf population.
They appeal to mechanisms either intrinsic to the galaxies themselves -- like a high-spin dark matter halo \citep{amorisco2016, rong2017, liao2019} or stellar feedback  \citep{dicintio2017, chan2018, jackson2020} -- or to the environment of the progenitors -- like ram pressure stripping \citep{yozin2015, tremmel2020, jackson2020}, tidal effects  \citep{carleton2019, jiang2019, liao2019}, or mergers  \citep{wright2020}.
These have focused primarily on explaining the necessary size increase to turn a canonical dwarf galaxy into an ``ultra-diffuse'' galaxy.

What we are most interested in, however, are the SFHs predicted or implied by these scenarios.
Stellar population gradients in particular provide significant insight into the SFH of a galaxy.
For instance, the stellar population gradients among the Local Group dSph population provide constraints for galaxy formation scenarios \citep[e.g.,][]{kawata2006, bl2016, elbadry2016, rj2018,  genina2019, mercado2020}.
For the canonical dwarf population, many formation scenarios take the positive age and negative metallicity gradients of the Local Group dSph population as reflective of an ``outside-in'' formation where star formation becomes increasingly centrally concentrated as a galaxy evolves \citep[e.g.,][]{bl2016, rj2018}.

In this scenario, dwarf galaxies with shorter, earlier SFHs will have flatter stellar population gradients than galaxies with ongoing SFHs, as is indicated by the purple arrow in Figure \ref{fig:gradient_context}. 
On the other hand, the FIRE simulations show that the gradients for the Local Group dwarf population could be reflective of an initial ``inside-out'' formation that was inverted due to stellar migration induced by stellar feedback \citep[e.g.,][orange arrow in Figure \ref{fig:gradient_context}]{elbadry2016}.

In either case, the flat age gradient and flat-to-positive metallicity gradient we measure for DF44 are contrary to expectations for an extended SFH, bursty or otherwise. 
Work using the FIRE-2 simulations shows that the diversity of metallicity gradients seen in the Local Group dSph population
(some that have strong gradients and others with comparatively weak gradients) 
can be explained by late, spatially-extended star-formation \citep[orange arrow in Fig \ref{fig:gradient_context};][]{mercado2020}. 
However, the two-component age we measure when fitting the high-S/N integrated spectrum in full-mode using \texttt{alf} does not show any evidence that DF44 experienced recent star-formation.

This is a crucial point because while \citet{dicintio2017} and \citet{chan2018} showed that internal stellar feedback from ongoing (protracted over many Gyr), bursty SFHs can puff up dwarf galaxies to make objects that look like UDGs, this mechanism cannot explain DF44.\footnote{The important caveat is that we are not reaching the outskirts of DF44 and we could potentially find declining age and metallicity profiles if we measured further out in galactocentric radius. However, because the observed color gradient does not change significantly past the reach of our stellar population gradients, we do not expect a significant change in age or metallicity.}

In addition, even the ``intrinsic'' mechanisms invoked to explain the sizes of UDGs still require environmental quenching to make the quiescent population of UDGs \citep{rong2017, liao2019, wright2020}.
We can assess the 
relevance of different UDG formation models for DF44 by their predictions for the infall times into high-density environments, 
and the corresponding epochs of quenching.

For DF44, any quenching related to infall would have had to happen early in order to agree with our measured old age.
This precludes scenarios that predict late infall/quenching including \citet{rong2017}, the field origin of group UDGs from \citet{liao2019}, and \citet{jackson2020}.

Several scenarios predict early quenching from ram pressure stripping \citep[e.g.,][]{yozin2015, tremmel2020} or tidal effects \citep[e.g.,][]{liao2019, carleton2019} after early infalls.
However, available phase-space information for some UDGs in the Coma cluster, including DF44, indicates that they are recent infalls \citep[$\lesssim 2$ Gyr;][]{alabi2018}. 
Moreover, \citet{vd2019} found that DF44 appears to be in a dynamically cold group with two other UDGs, DF42 and DFX2.

On the other hand, the radial alignment of the UDGs in Coma  \citep[i.e., the alignment of their major axis position angles aligned with the cluster center;][including DF44]{yagi2016} has been used as evidence of strong tidal interactions of the UDGs with the cluster \citep{carleton2019}.
There is also direct evidence for some UDGs having experienced significant tidal effects \citep{bennet2018, collins2020}.
However, by stacking images of all the Coma UDGs \citet{mowla2017} were able to rule out typical tidal features  down to an extremely faint limit ($\sim 34$ mag arcsec$^{-2}$; see their Figure 4).
Even if the individual UDGs have tidal features well below the detection threshold, they would have still manifested as a decrease in the ellipticity of the stack. 
If tidal features did exist as a general aspect of the Coma UDGs, they would have to be exactly aligned with the major axes of the UDGs which is not a phenomenon that has been observed elsewhere.

Moreover, since radial alignment requires significant time in the cluster, i.e., at least one pericenter passage, this scenario would require that the cold group is not actually real and that DF44's high orbital energy is not a result of recent infall but from being kicked out after an interaction with the cluster center.
This is not impossible, and it is difficult to conclusively rule out tidal effects from dynamics and imaging alone.
However, we note that in \citet{rong2020}, the  primordial spins and alignments seem preserved outside 0.5~$R_{200}$.
DF44 resides at 0.6~$R_{200}$, which is further evidence that it is unlikely to have 
had a significant interaction with the cluster.
Therefore, until more definitive evidence is presented otherwise, we conclude that that any morphological and star formation transformations DF44 experienced  happened before it was accreted into Coma.

In summary, none of the existing models for the origins of UDGs, or for dwarfs in general, appears to explain the properties of DF44. 
In the next subsection, we will outline what the observations to date do suggest about the evolutionary history of this galaxy.

\subsection{The Evolutionary History of DF44: Knowns and Unknowns}

We have found that the detailed kinematic and stellar population properties of DF44 are not consistent with those of canonical dwarf galaxies, nor is DF44 consistent with the various UDG formation scenarios proposed from simulations to date. 
That DF44 falls below the mass--metallicity relation is an important clue to its evidently exotic origins.
Generally, in the local Universe, relatively slow quenching mechanisms seem to be favored that allow for some vestigial star formation after the inflow of intergalactic gas has been cut off  \citep[e.g.,][]{peng2015, gallazzi2020arXiv}.
The comparatively low metallicity measured in DF44, its old age, and lack of evidence for an extended 
SFH\footnote{\citet{lee2020} found that DF44, along with most Coma UDGs, had significant star formation within the past Gyr, based on their ``blue'' ultraviolet-to-optical colors.
However, this work was based on expectations for metal-rich, massive galaxies, rather than for metal-poor dwarfs (e.g., \citealt{singh2019}), and in fact the colors appear to be fully consistent with an old population.}
all indicate that DF44 is a ``failed'' galaxy insofar as it seems to have quenched both early and catastrophically, with no vestigial star-formation. 
This early quenching is reminiscent of ultra-faint dwarfs \citep{simon2019}, but galaxies in the luminosity range of DF44 are not expected to quench without an environmental influence \citep[e.g.,][]{wheeler2019}, which is not present for DF44.

An internal quenching process for DF44 may be reflected in its stellar populations gradients.
We discussed in Section \ref{sec:comparison_to_observations} how the age and metallicity gradients for DF44 are unusual among similar mass but more compact galaxies.
Due to the difficulty in measuring granular differences in stellar ages for old populations \citep[see review by][]{conroy2013}, our measured age gradient could be much weaker than the true age difference.
Indeed, while our best-fit age gradient is flat, the uncertainties allow for it to potentially be either positive or negative.
Given that DF44 is old and there is no evidence for any recent star formation, we can also use [Mg/Fe] as a ``chemical clock,'' since in a given star-forming system, its value decreases with time as the interstellar medium is increasingly polluted by Type Ia supernovae.  
This means that the DF44 [Mg/Fe] gradient can be interpreted as a negative age gradient, and suggests that the galaxy lies in the upper left quadrant of Figure~\ref{fig:gradient_context} -- unlike what is currently measured for the canonical population of dwarfs.
Although the observational uncertainties are large, the increasing metallicity and decreasing $\alpha$-enhancement with radius suggest an inside-out scenario, where star-formation was cut off earlier in the center, i.e., the quenching process began there and moved outward, with more extended star formation toward the periphery of the galaxy.

Additionally, there is some theoretical basis in which a ``violent'' process such as mergers, rapid gas inflows, or strong feedback-driven outflows can both dilute the initial rotation in the system and flatten the internal metallicity gradient \citep[e.g.,][]{ho2015, ma2017}. 
This process may be closely linked with the populous GC system of DF44.
These compact objects must have formed in high-density regions of gas at an early epoch, quite unlike the present-day distribution of stars (see, e.g., \citealt{trujillogomez2020arXiv} for some basic physical considerations).  
The GC-like abundance patterns inside the central $\sim$~1~kpc of DF44 (Figure~\ref{fig:chemo_evo}) could reflect
this zone as the formation site of the GCs before they expanded outward (to the larger radii where they are observed today) along with the stars and dark matter.

The early quenching of DF44 (at $z \sim$~1--2 based on its inferred stellar age) would naturally lead to a present-day GC system that is overpopulous, and a stellar-to-halo mass ratio ($M_*/M_{\rm h}$) that is unusually low, relative to more typical dwarfs with similar stellar masses but later quenching.
Similar scenarios have previously been discussed
in the context of dwarfs with high GC specific frequencies that fall into clusters early  \citep{peng2008,liu2016,mistani2016, ra2020}, with recent work extending the concept to UDGs \citep{carleton_2020arXiv}.
However, DF44 was identified as a challenge to explain in this scenario, owing to its particularly high GC specific frequency, and here we also note that its {\it late} rather than early infall presents a fundamental distinction -- requiring a non-cluster quenching mechanism.

The high GC specific frequencies of some UDGs, including DF44, 
have been some of the key indicators of the failed-galaxy scenario
\citep{beasley2016a,peng2016,forbes2020}.
This is because GC numbers have been established as proxies for halo mass \citep[e.g.,][]{burkert2020},
which leads to the conclusion that the halos of these UDGs are overmassive (though not necessarily as extreme as Milky-Way mass).
This overall picture has been called into question in a reanalysis of DF44 by \citet{saifollahi2020}, who found much lower GC numbers than the previous work from \citet{vd2017b}.
The discrepancy between the two studies is puzzling, given their very similar methods applied to the same dataset, and may reflect the difficulties in separating GCs from contaminating sources on the outskirts of the galaxy.
More work on the GC system of DF44 is needed, and here we emphasize two points.
First is that the evidence for the failed-galaxy scenario from the stellar populations of DF44 is independent of this galaxy's precise GC number count.
Second is that {\it differential} analyses of Coma UDGs and normal dwarfs can circumvent the systematic uncertainties in establishing GC numbers in a single galaxy, and have shown that there is a subpopulation of UDGs with unusually populous GC systems
\citep{amorisco2018,lim2018}. 
Whether or not these UDGs also have unusual radial trends in their stellar populations, as in DF44, remains to be seen.

One missing piece to the DF44 puzzle is an unequivocal estimate of its halo mass, since the stellar dynamics constraints are still ambiguous.
In principle, diffuse gas in the halo (if such gas is present) can be used as a mass tracer, which has been attempted for DF44 in recent X-ray studies \citep{lee2020,bogdan2020}.
No signal was found, down to an X-ray luminosity of $L_X = 2\times10^{38}$~erg~s$^{-1}$, which was interpreted as evidence against a massive halo \citep{bogdan2020}\footnote{There
was also no detection of low-mass X-ray binaries (LMXBs), which are preferentially formed in GCs.
This was interpreted as indirect evidence against large numbers of GCs in DF44, but the well-known connections between LMXBs and the metallicities and sizes of star clusters (e.g., \citealt{sivakoff2007,hou2016}) were not examined in detail, and could plausibly explain the non-detection.}.
However, this conclusion was premature, as it used an empirical relation between $L_X$ and aperture mass that was tied to the sizes of giant elliptical galaxies rather than to UDGs.
It is straightforward to find examples of nearby Milky-Way-mass galaxies with $L_X$ below the detection limit at Coma-cluster distances 
\citep[e.g., NGC 3377;][]{kim2013, alabi2017},
and the general prospects for obtaining usefully strong direct constraints on the DF44 halo mass remain unclear.

While DF44 is the most thoroughly studied UDG to date, it is not the only example of the failed-galaxy sub-type, nor the most extreme.
Some UDGs in Coma have GC systems comprising $\sim$~10\% of their stellar mass \citep{forbes2020}, and could represent ``pure stellar halos'' (since the bulk of the field stars could have originated from coeval clusters that have dissolved or shed substantial mass).
A menagerie of quiescent UDGs in low-density environments may well belong to the same family \citep{papastergis2017,alabi2018}.
Here stellar populations is an emerging, powerful avenue for identifying and characterizing failed galaxies.
NGC~5846\_UDG1 is a relatively nearby galaxy in a group, 
with little rotation and an overpopulous GC system \citep{forbes2020arXiv},
where the field stars and GCs were found to have the same age and metallicity \citep{muller2020} --
in contrast to typical galaxies (e.g., \citealt{larsen2014a}).
DGSAT~I is a more isolated UDG with an extreme metal-poor, $\alpha$-enhanced abundance pattern that suggests a primordial galaxy \citep{martin-navarro2019}.
Further deep spectroscopy of other UDGs and LSB galaxies will be key to understanding their nature and origins.

How the exotic formation histories of UDGs such as DF44 connect to standard models of galaxy formation and evolution is unclear.
As discussed, these galaxies may have novel modes of internal feedback that are related to GC formation, while relatively obscure environmental effects such as cosmic web stripping may also be relevant \citep{bl2013}.
On the other hand, the UDGs may yield broader implications for galaxy evolution -- particularly for dwarf galaxies which 
should be the easiest to simulate in detail, yet harbor various unsolved problems.
There have been extensive efforts to understand the dark matter distributions in dwarfs, with general agreement now that previous puzzles could be solved naturally with standard processes including baryonic feedback (e.g., \citealt{bullock2017}).
However, there are still problems in reproducing their basic baryonic properties that can be directly observed 
(e.g., the broad distribution of sizes in dwarfs, and the metallicities of ``primordial'' ultra-faint galaxies;
\citealt{jiang2019,wheeler2019}).
Galaxy evolution in general remains a major unsolved problem, and the same physical
processes that still need solving for UDGs may be relevant for unsolved
problems more generally.

\section{Summary}\label{s.df44.summary}

In this work we applied SPS models to the optical spectra of the UDG DF44 using the IFU instrument KCWI on the Keck II telescope. The huge gain in S/N compared to other stellar population studies of UDGs allowed us to significantly reduce the uncertainties of the global stellar population measurements and, for the first time for a UDG, present spatially-resolved stellar population measurements. 
This approach provides a new avenue for helping to discriminate between UDG sub-types and to clarify their evolutionary histories.

We summarize our findings as follow:

\begin{itemize}
    \item DF44 represents an extreme of the UDG population; old and metal-poor to the point that it falls below the mass--metallicity relation established by the canonical dwarf population. 
    A two-age fit to the integrated spectrum shows no evidence for a young stellar component.
    We infer that DF44 quenched early and catastrophically, prior to infall into the Coma cluster. 
    
    \item  The gradients of the spatially-resolved diffuse-stellar-light parameters are as follows:
      \ageslope log dex kpc$^{-1}$, \fehslope dex kpc$^{-1}$, and \mgfslope dex kpc$^{-1}$.
     Star formation apparently proceeded in an ``inside-out'' manner within DF44, with the central regions potentially coeval with the GC population.
    These abundance patterns contrast with fairly universal expectations that dwarfs have more extended star formation histories in their centers.
    
    \item The stellar population results, along with the previously published kinematic results, indicate that DF44 is dissimilar to canonical dwarf populations.
    This galaxy, as well as other UDGs with unusual properties, is in tension
    with current formation models wherein UDGs are contiguous with the normal dwarf distribution evolving under the influence of conventional processes.
\end{itemize}

Further insight into the formation histories of UDGs, and of dwarf galaxies more generally, 
should come through systematic, deep spectroscopy of more objects.
Fresh theoretical work is also critically needed, to move beyond general reproduction of size--mass trends and focus on their GC systems
(e.g., \citealt{carleton_2020arXiv,trujillogomez2020arXiv}) and, as is now possible, their detailed internal properties.

\software{IPython \citep{PER-GRA:2007}, 
          SciPy \citep{2020SciPy-NMeth},
          NumPy  \citep{van2011numpy}, 
          matplotlib  \citep{Hunter:2007},
          Astropy \citep{astropy18},
          PyMC3 \citep{pymc3},
          SPI \citep{villaume2017a}
          alf \citep{conroy2018a},
          EZ\_Ages \citep{schiavon2007, graves2008}
}

\acknowledgments
We would like to thank the anonymous referee for their constructive comments, J. Gannon for some discussion about GC candidate selection, M. Gu for help with various technical odds and ends, S. Laine for discussions about SFHs, R. Schiavon for helping with \texttt{EZ\_Ages}, and J. Taylor for very helpful discussion about the dynamics of satellites in clusters.  This research was supported in part by the National Science Foundation under Grant No. NSF PHY-1748958 through the Kavli Institute for Theoretical Physics workshop Globular Clusters at the Nexus of Star and Galaxy Formation for enabling useful discussions relevant to this paper. 
A.V. would like to acknowledge the NSF Graduate Fellowship, the UC Santa Cruz Chancellor's Dissertation Year Fellowship, and the Waterloo Centre Astrophysics Postdoctoral Fellowship for their support.
A.J.R. was supported by National Science Foundation grant AST-1616710, and as a Research Corporation for Science Advancement Cottrell Scholar.
S.D. is supported by NASA through Hubble Fellowship grant \#HST-HF2-51454.001-A awarded by the Space Telescope Science Institute, which is operated by the Association of Universities for Research in Astronomy, Incorporated, under NASA contract NAS5-26555.

\bibliography{references}{}

\begin{thebibliography}{}
\expandafter\ifx\csname natexlab\endcsname\relax\def\natexlab#1{#1}\fi
\providecommand{\url}[1]{\href{#1}{#1}}
\providecommand{\dodoi}[1]{doi:~\href{http://doi.org/#1}{\nolinkurl{#1}}}
\providecommand{\doeprint}[1]{\href{http://ascl.net/#1}{\nolinkurl{http://ascl.net/#1}}}
\providecommand{\doarXiv}[1]{\href{https://arxiv.org/abs/#1}{\nolinkurl{https://arxiv.org/abs/#1}}}

\bibitem[{{Abraham} \& {van Dokkum}(2014)}]{abraham2014}
{Abraham}, R.~G., \& {van Dokkum}, P.~G. 2014, \pasp, 126, 55,
  \dodoi{10.1086/674875}

\bibitem[{{Alabi} {et~al.}(2018){Alabi}, {Ferr{\'e}-Mateu}, {Romanowsky},
  {Brodie}, {Forbes}, {Wasserman}, {Bellstedt}, {Mart{\'\i}n-Navarro},
  {Pandya}, {Stone}, \& {Okabe}}]{alabi2018}
{Alabi}, A., {Ferr{\'e}-Mateu}, A., {Romanowsky}, A.~J., {et~al.} 2018, \mnras,
  479, 3308, \dodoi{10.1093/mnras/sty1616}

\bibitem[{{Alabi} {et~al.}(2017){Alabi}, {Forbes}, {Romanowsky}, {Brodie},
  {Strader}, {Janz}, {Usher}, {Spitler}, {Bellstedt}, \&
  {Ferr{\'e}-Mateu}}]{alabi2017}
{Alabi}, A.~B., {Forbes}, D.~A., {Romanowsky}, A.~J., {et~al.} 2017, \mnras,
  468, 3949, \dodoi{10.1093/mnras/stx678}

\bibitem[{{Albers} {et~al.}(2019){Albers}, {Weisz}, {Cole}, {Dolphin},
  {Skillman}, {Williams}, {Boylan-Kolchin}, {Bullock}, {Dalcanton}, {Hopkins},
  {Leaman}, {McConnachie}, {Vogelsberger}, \& {Wetzel}}]{albers2019}
{Albers}, S.~M., {Weisz}, D.~R., {Cole}, A.~A., {et~al.} 2019, \mnras, 490,
  5538, \dodoi{10.1093/mnras/stz2903}

\bibitem[{{Amorisco} \& {Loeb}(2016)}]{amorisco2016}
{Amorisco}, N.~C., \& {Loeb}, A. 2016, \mnras, 459, L51,
  \dodoi{10.1093/mnrasl/slw055}

\bibitem[{{Amorisco} {et~al.}(2018){Amorisco}, {Monachesi}, {Agnello}, \&
  {White}}]{amorisco2018}
{Amorisco}, N.~C., {Monachesi}, A., {Agnello}, A., \& {White}, S.~D.~M. 2018,
  \mnras, 475, 4235, \dodoi{10.1093/mnras/sty116}

\bibitem[{{Astropy Collaboration} {et~al.}(2018){Astropy Collaboration},
  {Price-Whelan}, {Sip{H o}cz}, {G{"u}nther}, {Lim}, {Crawford}, {Conseil},
  {Shupe}, {Craig}, {Dencheva}, {Ginsburg}, {VanderPlas}, {Bradley},
  {P{'e}rez-Su{'a}rez}, {de Val-Borro}, {Aldcroft}, {Cruz}, {Robitaille},
  {Tollerud}, {Ardelean}, {Babej}, {Bach}, {Bachetti}, {Bakanov}, {Bamford},
  {Barentsen}, {Barmby}, {Baumbach}, {Berry}, {Biscani}, {Boquien}, {Bostroem},
  {Bouma}, {Brammer}, {Bray}, {Breytenbach}, {Buddelmeijer}, {Burke},
  {Calderone}, {Cano Rodr{'{i}}guez}, {Cara}, {Cardoso}, {Cheedella}, {Copin},
  {Corrales}, {Crichton}, {D'Avella}, {Deil}, {Depagne}, {Dietrich}, {Donath},
  {Droettboom}, {Earl}, {Erben}, {Fabbro}, {Ferreira}, {Finethy}, {Fox},
  {Garrison}, {Gibbons}, {Goldstein}, {Gommers}, {Greco}, {Greenfield},
  {Groener}, {Grollier}, {Hagen}, {Hirst}, {Homeier}, {Horton}, {Hosseinzadeh},
  {Hu}, {Hunkeler}, {Ivezi{'c}}, {Jain}, {Jenness}, {Kanarek}, {Kendrew},
  {Kern}, {Kerzendorf}, {Khvalko}, {King}, {Kirkby}, {Kulkarni}, {Kumar},
  {Lee}, {Lenz}, {Littlefair}, {Ma}, {Macleod}, {Mastropietro}, {McCully},
  {Montagnac}, {Morris}, {Mueller}, {Mumford}, {Muna}, {Murphy}, {Nelson},
  {Nguyen}, {Ninan}, {N{"o}the}, {Ogaz}, {Oh}, {Parejko}, {Parley}, {Pascual},
  {Patil}, {Patil}, {Plunkett}, {Prochaska}, {Rastogi}, {Reddy Janga},
  {Sabater}, {Sakurikar}, {Seifert}, {Sherbert}, {Sherwood-Taylor}, {Shih},
  {Sick}, {Silbiger}, {Singanamalla}, {Singer}, {Sladen}, {Sooley},
  {Sornarajah}, {Streicher}, {Teuben}, {Thomas}, {Tremblay}, {Turner},
  {Terr{'o}n}, {van Kerkwijk}, {de la Vega}, {Watkins}, {Weaver}, {Whitmore},
  {Woillez}, {Zabalza}, \& {Astropy Contributors}}]{astropy18}
{Astropy Collaboration}, {Price-Whelan}, A.~M., {Sip{H o}cz}, B.~M., {et~al.}
  2018, aj, 156, 123, \dodoi{10.3847/1538-3881/aabc4f}

\bibitem[{{Beasley} {et~al.}(2016){Beasley}, {Romanowsky}, {Pota}, {Navarro},
  {Martinez Delgado}, {Neyer}, \& {Deich}}]{beasley2016a}
{Beasley}, M.~A., {Romanowsky}, A.~J., {Pota}, V., {et~al.} 2016, \apjl, 819,
  L20, \dodoi{10.3847/2041-8205/819/2/L20}

\bibitem[{{Bellazzini} {et~al.}(2017){Bellazzini}, {Belokurov}, {Magrini},
  {Fraternali}, {Testa}, {Beccari}, {Marchetti}, \& {Carini}}]{bellazzini2017}
{Bellazzini}, M., {Belokurov}, V., {Magrini}, L., {et~al.} 2017, \mnras, 467,
  3751, \dodoi{10.1093/mnras/stx236}

\bibitem[{{Ben{\'\i}tez-Llambay} {et~al.}(2013){Ben{\'\i}tez-Llambay},
  {Navarro}, {Abadi}, {Gottl{\"o}ber}, {Yepes}, {Hoffman}, \&
  {Steinmetz}}]{bl2013}
{Ben{\'\i}tez-Llambay}, A., {Navarro}, J.~F., {Abadi}, M.~G., {et~al.} 2013,
  \apjl, 763, L41, \dodoi{10.1088/2041-8205/763/2/L41}

\bibitem[{{Ben{\'\i}tez-Llambay} {et~al.}(2016){Ben{\'\i}tez-Llambay},
  {Navarro}, {Abadi}, {Gottl{\"o}ber}, {Yepes}, {Hoffman}, \&
  {Steinmetz}}]{bl2016}
---. 2016, \mnras, 456, 1185, \dodoi{10.1093/mnras/stv2722}

\bibitem[{{Bennet} {et~al.}(2018){Bennet}, {Sand}, {Zaritsky}, {Crnojevi{\'c}},
  {Spekkens}, \& {Karunakaran}}]{bennet2018}
{Bennet}, P., {Sand}, D.~J., {Zaritsky}, D., {et~al.} 2018, \apjl, 866, L11,
  \dodoi{10.3847/2041-8213/aadedf}

\bibitem[{{Bogd{\'a}n}(2020)}]{bogdan2020}
{Bogd{\'a}n}, {\'A}. 2020, \apjl, 901, L30, \dodoi{10.3847/2041-8213/abb886}

\bibitem[{{Bothun} {et~al.}(1991){Bothun}, {Impey}, \& {Malin}}]{bothun1991}
{Bothun}, G.~D., {Impey}, C.~D., \& {Malin}, D.~F. 1991, \apj, 376, 404,
  \dodoi{10.1086/170290}

\bibitem[{{Bullock} \& {Boylan-Kolchin}(2017)}]{bullock2017}
{Bullock}, J.~S., \& {Boylan-Kolchin}, M. 2017, \araa, 55, 343,
  \dodoi{10.1146/annurev-astro-091916-055313}

\bibitem[{{Burkert} \& {Forbes}(2020)}]{burkert2020}
{Burkert}, A., \& {Forbes}, D.~A. 2020, \aj, 159, 56,
  \dodoi{10.3847/1538-3881/ab5b0e}

\bibitem[{{Carleton} {et~al.}(2019){Carleton}, {Errani}, {Cooper},
  {Kaplinghat}, {Pe{\~n}arrubia}, \& {Guo}}]{carleton2019}
{Carleton}, T., {Errani}, R., {Cooper}, M., {et~al.} 2019, \mnras, 485, 382,
  \dodoi{10.1093/mnras/stz383}

\bibitem[{{Carleton} {et~al.}(2020){Carleton}, {Guo}, {Munshi}, {Tremmel}, \&
  {Wright}}]{carleton_2020arXiv}
{Carleton}, T., {Guo}, Y., {Munshi}, F., {Tremmel}, M., \& {Wright}, A. 2020,
  arXiv e-prints, arXiv:2008.11205.
\newblock \doarXiv{2008.11205}

\bibitem[{{Chan} {et~al.}(2018){Chan}, {Kere{\v{s}}}, {Wetzel}, {Hopkins},
  {Faucher-Gigu{\`e}re}, {El-Badry}, {Garrison-Kimmel}, \&
  {Boylan-Kolchin}}]{chan2018}
{Chan}, T.~K., {Kere{\v{s}}}, D., {Wetzel}, A., {et~al.} 2018, \mnras, 478,
  906, \dodoi{10.1093/mnras/sty1153}

\bibitem[{{Chilingarian} {et~al.}(2019){Chilingarian}, {Afanasiev}, {Grishin},
  {Fabricant}, \& {Moran}}]{chilingarian2019}
{Chilingarian}, I.~V., {Afanasiev}, A.~V., {Grishin}, K.~A., {Fabricant}, D.,
  \& {Moran}, S. 2019, \apj, 884, 79, \dodoi{10.3847/1538-4357/ab4205}

\bibitem[{{Choi} {et~al.}(2014){Choi}, {Conroy}, {Moustakas}, {Graves},
  {Holden}, {Brodwin}, {Brown}, \& {van Dokkum}}]{choi2014}
{Choi}, J., {Conroy}, C., {Moustakas}, J., {et~al.} 2014, \apj, 792, 95,
  \dodoi{10.1088/0004-637X/792/2/95}

\bibitem[{{Choi} {et~al.}(2016){Choi}, {Dotter}, {Conroy}, {Cantiello},
  {Paxton}, \& {Johnson}}]{choi2016}
{Choi}, J., {Dotter}, A., {Conroy}, C., {et~al.} 2016, \apj, 823, 102,
  \dodoi{10.3847/0004-637X/823/2/102}

\bibitem[{{Collins} {et~al.}(2020){Collins}, {Tollerud}, {Rich}, {Ibata},
  {Martin}, {Chapman}, {Gilbert}, \& {Preston}}]{collins2020}
{Collins}, M. L.~M., {Tollerud}, E.~J., {Rich}, R.~M., {et~al.} 2020, \mnras,
  491, 3496, \dodoi{10.1093/mnras/stz3252}

\bibitem[{{Conroy}(2013)}]{conroy2013}
{Conroy}, C. 2013, \araa, 51, 393, \dodoi{10.1146/annurev-astro-082812-141017}

\bibitem[{{Conroy} \& {Gunn}(2010)}]{conroy2010}
{Conroy}, C., \& {Gunn}, J.~E. 2010, \apj, 712, 833,
  \dodoi{10.1088/0004-637X/712/2/833}

\bibitem[{{Conroy} {et~al.}(2018){Conroy}, {Villaume}, {van Dokkum}, \&
  {Lind}}]{conroy2018a}
{Conroy}, C., {Villaume}, A., {van Dokkum}, P.~G., \& {Lind}, K. 2018, \apj,
  854, 139, \dodoi{10.3847/1538-4357/aaab49}

\bibitem[{{Dalcanton} {et~al.}(1997){Dalcanton}, {Spergel}, {Gunn}, {Schmidt},
  \& {Schneider}}]{dalcanton1997}
{Dalcanton}, J.~J., {Spergel}, D.~N., {Gunn}, J.~E., {Schmidt}, M., \&
  {Schneider}, D.~P. 1997, \aj, 114, 635, \dodoi{10.1086/118499}

\bibitem[{{Danieli} {et~al.}(2019){Danieli}, {van Dokkum}, {Conroy}, {Abraham},
  \& {Romanowsky}}]{danieli2019}
{Danieli}, S., {van Dokkum}, P., {Conroy}, C., {Abraham}, R., \& {Romanowsky},
  A.~J. 2019, \apjl, 874, L12, \dodoi{10.3847/2041-8213/ab0e8c}

\bibitem[{{Di Cintio} {et~al.}(2017){Di Cintio}, {Brook}, {Dutton},
  {Macci{\`o}}, {Obreja}, \& {Dekel}}]{dicintio2017}
{Di Cintio}, A., {Brook}, C.~B., {Dutton}, A.~A., {et~al.} 2017, \mnras, 466,
  L1, \dodoi{10.1093/mnrasl/slw210}

\bibitem[{{El-Badry} {et~al.}(2016){El-Badry}, {Wetzel}, {Geha}, {Hopkins},
  {Kere{\v{s}}}, {Chan}, \& {Faucher-Gigu{\`e}re}}]{elbadry2016}
{El-Badry}, K., {Wetzel}, A., {Geha}, M., {et~al.} 2016, \apj, 820, 131,
  \dodoi{10.3847/0004-637X/820/2/131}

\bibitem[{{Faria} {et~al.}(2007){Faria}, {Feltzing}, {Lundstr{\"o}m},
  {Gilmore}, {Wahlgren}, {Ardeberg}, \& {Linde}}]{faria2007}
{Faria}, D., {Feltzing}, S., {Lundstr{\"o}m}, I., {et~al.} 2007, \aap, 465,
  357, \dodoi{10.1051/0004-6361:20065244}

\bibitem[{{Ferr{\'e}-Mateu} {et~al.}(2018){Ferr{\'e}-Mateu}, {Alabi}, {Forbes},
  {Romanowsky}, {Brodie}, {Pandya}, {Mart{\'\i}n-Navarro}, {Bellstedt},
  {Wasserman}, {Stone}, \& {Okabe}}]{ferre-mateu2018}
{Ferr{\'e}-Mateu}, A., {Alabi}, A., {Forbes}, D.~A., {et~al.} 2018, \mnras,
  479, 4891, \dodoi{10.1093/mnras/sty1597}

\bibitem[{{Forbes} {et~al.}(2020{\natexlab{a}}){Forbes}, {Alabi}, {Romanowsky},
  {Brodie}, \& {Arimoto}}]{forbes2020}
{Forbes}, D.~A., {Alabi}, A., {Romanowsky}, A.~J., {Brodie}, J.~P., \&
  {Arimoto}, N. 2020{\natexlab{a}}, \mnras, 175, \dodoi{10.1093/mnras/staa180}

\bibitem[{{Forbes} {et~al.}(2020{\natexlab{b}}){Forbes}, {Gannon},
  {Romanowsky}, {Alabi}, {Brodie}, {Couch}, \& {Ferre-Mateu}}]{forbes2020arXiv}
{Forbes}, D.~A., {Gannon}, J.~S., {Romanowsky}, A.~J., {et~al.}
  2020{\natexlab{b}}, arXiv e-prints, arXiv:2010.07313.
\newblock \doarXiv{2010.07313}

\bibitem[{{Foreman-Mackey} {et~al.}(2013){Foreman-Mackey}, {Hogg}, {Lang}, \&
  {Goodman}}]{emcee_v1}
{Foreman-Mackey}, D., {Hogg}, D.~W., {Lang}, D., \& {Goodman}, J. 2013, \pasp,
  125, 306, \dodoi{10.1086/670067}

\bibitem[{{Foreman-Mackey} {et~al.}(2014){Foreman-Mackey}, {Sick}, \&
  {Johnson}}]{pythonfsps}
{Foreman-Mackey}, D., {Sick}, J., \& {Johnson}, B. 2014, {Python-Fsps: Python
  Bindings To Fsps (V0.1.1)}, v0.1.1,  Zenodo, \dodoi{10.5281/zenodo.12157}

\bibitem[{{Gallazzi} {et~al.}(2020){Gallazzi}, {Pasquali}, {Zibetti}, \& {La
  Barbera}}]{gallazzi2020arXiv}
{Gallazzi}, A.~R., {Pasquali}, A., {Zibetti}, S., \& {La Barbera}, F. 2020,
  arXiv e-prints, arXiv:2010.04733.
\newblock \doarXiv{2010.04733}

\bibitem[{{Gannon} {et~al.}(2020){Gannon}, {Forbes}, {Romanowsky},
  {Ferr{\'e}-Mateu}, {Couch}, \& {Brodie}}]{gannon2020}
{Gannon}, J.~S., {Forbes}, D.~A., {Romanowsky}, A.~J., {et~al.} 2020, \mnras,
  495, 2582, \dodoi{10.1093/mnras/staa1282}

\bibitem[{{Geha} {et~al.}(2010){Geha}, {van der Marel}, {Guhathakurta},
  {Gilbert}, {Kalirai}, \& {Kirby}}]{geha2010}
{Geha}, M., {van der Marel}, R.~P., {Guhathakurta}, P., {et~al.} 2010, \apj,
  711, 361, \dodoi{10.1088/0004-637X/711/1/361}

\bibitem[{{Genina} {et~al.}(2019){Genina}, {Frenk}, {Ben{\'\i}tez-Llambay},
  {Cole}, {Navarro}, {Oman}, \& {Fattahi}}]{genina2019}
{Genina}, A., {Frenk}, C.~S., {Ben{\'\i}tez-Llambay}, A.~r., {et~al.} 2019,
  \mnras, 488, 2312, \dodoi{10.1093/mnras/stz1852}

\bibitem[{{Graves} \& {Schiavon}(2008)}]{graves2008}
{Graves}, G.~J., \& {Schiavon}, R.~P. 2008, \apjs, 177, 446,
  \dodoi{10.1086/588097}

\bibitem[{{Gu} {et~al.}(2018){Gu}, {Conroy}, {Law}, {van Dokkum}, {Yan},
  {Wake}, {Bundy}, {Merritt}, {Abraham}, {Zhang}, {Bershady}, {Bizyaev},
  {Brinkmann}, {Drory}, {Grabowski}, {Masters}, {Pan}, {Parejko}, {Weijmans},
  \& {Zhang}}]{gu2018a}
{Gu}, M., {Conroy}, C., {Law}, D., {et~al.} 2018, \apj, 859, 37,
  \dodoi{10.3847/1538-4357/aabbae}

\bibitem[{{Harbeck} {et~al.}(2001){Harbeck}, {Grebel}, {Holtzman},
  {Guhathakurta}, {Brandner}, {Geisler}, {Sarajedini}, {Dolphin},
  {Hurley-Keller}, \& {Mateo}}]{harbeck2001}
{Harbeck}, D., {Grebel}, E.~K., {Holtzman}, J., {et~al.} 2001, \aj, 122, 3092,
  \dodoi{10.1086/324232}

\bibitem[{{Hayes} {et~al.}(2020){Hayes}, {Majewski}, {Hasselquist}, {Anguiano},
  {Shetrone}, {Law}, {Schiavon}, {Cunha}, {Smith}, {Beaton}, {Price-Whelan},
  {Allende Prieto}, {Battaglia}, {Bizyaev}, {Brownstein}, {Cohen},
  {Frinchaboy}, {Garc{\'\i}a-Hern{\'a}ndez}, {Lacerna}, {Lane},
  {M{\'e}sz{\'a}ros}, {Bidin}, {M{\~{u}}noz}, {Nidever}, {Oravetz}, {Oravetz},
  {Pan}, {Roman-Lopes}, {Sobeck}, \& {Stringfellow}}]{hayes2020}
{Hayes}, C.~R., {Majewski}, S.~R., {Hasselquist}, S., {et~al.} 2020, \apj, 889,
  63, \dodoi{10.3847/1538-4357/ab62ad}

\bibitem[{{Ho} {et~al.}(2015){Ho}, {Kudritzki}, {Kewley}, {Zahid}, {Dopita},
  {Bresolin}, \& {Rupke}}]{ho2015}
{Ho}, I.~T., {Kudritzki}, R.-P., {Kewley}, L.~J., {et~al.} 2015, \mnras, 448,
  2030, \dodoi{10.1093/mnras/stv067}

\bibitem[{{Hopkins} {et~al.}(2018){Hopkins}, {Wetzel}, {Kere{\v{s}}},
  {Faucher-Gigu{\`e}re}, {Quataert}, {Boylan-Kolchin}, {Murray}, {Hayward},
  {Garrison-Kimmel}, {Hummels}, {Feldmann}, {Torrey}, {Ma},
  {Angl{\'e}s-Alc{\'a}zar}, {Su}, {Orr}, {Schmitz}, {Escala}, {Sanderson},
  {Grudi{\'c}}, {Hafen}, {Kim}, {Fitts}, {Bullock}, {Wheeler}, {Chan},
  {Elbert}, \& {Narayanan}}]{hopkins2018}
{Hopkins}, P.~F., {Wetzel}, A., {Kere{\v{s}}}, D., {et~al.} 2018, \mnras, 480,
  800, \dodoi{10.1093/mnras/sty1690}

\bibitem[{{Hou} \& {Li}(2016)}]{hou2016}
{Hou}, M., \& {Li}, Z. 2016, \apj, 819, 164,
  \dodoi{10.3847/0004-637X/819/2/164}

\bibitem[{Hunter(2007)}]{Hunter:2007}
Hunter, J.~D. 2007, Computing In Science \& Engineering, 9, 90

\bibitem[{{Impey} {et~al.}(1988){Impey}, {Bothun}, \& {Malin}}]{impey1988}
{Impey}, C., {Bothun}, G., \& {Malin}, D. 1988, \apj, 330, 634,
  \dodoi{10.1086/166500}

\bibitem[{{Jackson} {et~al.}(2020){Jackson}, {Martin}, {Kaviraj}, {Rams{\o}y},
  {Devriendt}, {Sedgwick}, {Laigle}, {Choi}, {Beckmann}, {Volonteri}, {Dubois},
  {Pichon}, {Yi}, {Slyz}, {Kraljic}, {Kimm}, {Peirani}, \&
  {Baldry}}]{jackson2020}
{Jackson}, R.~A., {Martin}, G., {Kaviraj}, S., {et~al.} 2020, arXiv e-prints,
  arXiv:2007.06581.
\newblock \doarXiv{2007.06581}

\bibitem[{{Jiang} {et~al.}(2019){Jiang}, {Dekel}, {Freundlich}, {Romanowsky},
  {Dutton}, {Macci{\`o}}, \& {Di Cintio}}]{jiang2019}
{Jiang}, F., {Dekel}, A., {Freundlich}, J., {et~al.} 2019, \mnras, 487, 5272,
  \dodoi{10.1093/mnras/stz1499}

\bibitem[{{Kacharov} {et~al.}(2017){Kacharov}, {Battaglia}, {Rejkuba}, {Cole},
  {Carrera}, {Fraternali}, {Wilkinson}, {Gallart}, {Irwin}, \&
  {Tolstoy}}]{kacharov2017}
{Kacharov}, N., {Battaglia}, G., {Rejkuba}, M., {et~al.} 2017, \mnras, 466,
  2006, \dodoi{10.1093/mnras/stw3188}

\bibitem[{{Kawata} {et~al.}(2006){Kawata}, {Arimoto}, {Cen}, \&
  {Gibson}}]{kawata2006}
{Kawata}, D., {Arimoto}, N., {Cen}, R., \& {Gibson}, B.~K. 2006, \apj, 641,
  785, \dodoi{10.1086/500633}

\bibitem[{{Kim} \& {Fabbiano}(2013)}]{kim2013}
{Kim}, D.-W., \& {Fabbiano}, G. 2013, \apj, 776, 116,
  \dodoi{10.1088/0004-637X/776/2/116}

\bibitem[{{Kirby} {et~al.}(2013){Kirby}, {Cohen}, {Guhathakurta}, {Cheng},
  {Bullock}, \& {Gallazzi}}]{kirby2013}
{Kirby}, E.~N., {Cohen}, J.~G., {Guhathakurta}, P., {et~al.} 2013, \apj, 779,
  102, \dodoi{10.1088/0004-637X/779/2/102}

\bibitem[{{Kirby} {et~al.}(2009){Kirby}, {Guhathakurta}, {Bolte}, {Sneden}, \&
  {Geha}}]{kirby2009}
{Kirby}, E.~N., {Guhathakurta}, P., {Bolte}, M., {Sneden}, C., \& {Geha}, M.~C.
  2009, \apj, 705, 328, \dodoi{10.1088/0004-637X/705/1/328}

\bibitem[{{Koda} {et~al.}(2015){Koda}, {Yagi}, {Yamanoi}, \&
  {Komiyama}}]{koda2015}
{Koda}, J., {Yagi}, M., {Yamanoi}, H., \& {Komiyama}, Y. 2015, \apjl, 807, L2,
  \dodoi{10.1088/2041-8205/807/1/L2}

\bibitem[{{Koleva} {et~al.}(2011){Koleva}, {Prugniel}, {De Rijcke}, \&
  {Zeilinger}}]{koleva2011}
{Koleva}, M., {Prugniel}, P., {De Rijcke}, S., \& {Zeilinger}, W.~W. 2011,
  \mnras, 417, 1643, \dodoi{10.1111/j.1365-2966.2011.19057.x}

\bibitem[{{Kroupa}(2001)}]{kroupa2001}
{Kroupa}, P. 2001, \mnras, 322, 231, \dodoi{10.1046/j.1365-8711.2001.04022.x}

\bibitem[{{Larsen} {et~al.}(2014){Larsen}, {Brodie}, {Forbes}, \&
  {Strader}}]{larsen2014a}
{Larsen}, S.~S., {Brodie}, J.~P., {Forbes}, D.~A., \& {Strader}, J. 2014, \aap,
  565, A98, \dodoi{10.1051/0004-6361/201322672}

\bibitem[{{Leaman} {et~al.}(2013){Leaman}, {Venn}, {Brooks}, {Battaglia},
  {Cole}, {Ibata}, {Irwin}, {McConnachie}, {Mendel}, {Starkenburg}, \&
  {Tolstoy}}]{leaman2013}
{Leaman}, R., {Venn}, K.~A., {Brooks}, A.~M., {et~al.} 2013, \apj, 767, 131,
  \dodoi{10.1088/0004-637X/767/2/131}

\bibitem[{{Lee} {et~al.}(2020){Lee}, {Hodges-Kluck}, \& {Gallo}}]{lee2020}
{Lee}, C.~H., {Hodges-Kluck}, E., \& {Gallo}, E. 2020, \mnras, 497, 2759,
  \dodoi{10.1093/mnras/staa1955}

\bibitem[{{Liao} {et~al.}(2019){Liao}, {Gao}, {Frenk}, {Grand}, {Guo},
  {G{\'o}mez}, {Marinacci}, {Pakmor}, {Shao}, \& {Springel}}]{liao2019}
{Liao}, S., {Gao}, L., {Frenk}, C.~S., {et~al.} 2019, \mnras, 490, 5182,
  \dodoi{10.1093/mnras/stz2969}

\bibitem[{{Lim} {et~al.}(2018){Lim}, {Peng}, {C{\^o}t{\'e}}, {Sales}, {den
  Brok}, {Blakeslee}, \& {Guhathakurta}}]{lim2018}
{Lim}, S., {Peng}, E.~W., {C{\^o}t{\'e}}, P., {et~al.} 2018, \apj, 862, 82,
  \dodoi{10.3847/1538-4357/aacb81}

\bibitem[{{Liu} \& {Graham}(2001)}]{liu2001}
{Liu}, M.~C., \& {Graham}, J.~R. 2001, \apjl, 557, L31, \dodoi{10.1086/323174}

\bibitem[{{Liu} {et~al.}(2016){Liu}, {Peng}, {Blakeslee}, {C{\^o}t{\'e}},
  {Ferrarese}, {Jord{\'a}n}, {Puzia}, {Toloba}, \& {Zhang}}]{liu2016}
{Liu}, Y., {Peng}, E.~W., {Blakeslee}, J., {et~al.} 2016, \apj, 818, 179,
  \dodoi{10.3847/0004-637X/818/2/179}

\bibitem[{{Longobardi} {et~al.}(2018){Longobardi}, {Peng}, {C{\^o}t{\'e}},
  {Mihos}, {Ferrarese}, {Puzia}, {Lan{\c{c}}on}, {Zhang}, {Mu{\~n}oz},
  {Blakeslee}, {Guhathakurta}, {Durrell}, {S{\'a}nchez-Janssen}, {Toloba},
  {Jord{\'a}n}, {Eyheramendy}, {Cuilland re}, {Gwyn}, {Boselli}, {Duc}, {Liu},
  {Alamo-Mart{\'\i}nez}, {Powalka}, \& {Lim}}]{longobardi2018a}
{Longobardi}, A., {Peng}, E.~W., {C{\^o}t{\'e}}, P., {et~al.} 2018, \apj, 864,
  36, \dodoi{10.3847/1538-4357/aad3d2}

\bibitem[{{Ma} {et~al.}(2017){Ma}, {Hopkins}, {Feldmann}, {Torrey},
  {Faucher-Gigu{\`e}re}, \& {Kere{\v{s}}}}]{ma2017}
{Ma}, X., {Hopkins}, P.~F., {Feldmann}, R., {et~al.} 2017, \mnras, 466, 4780,
  \dodoi{10.1093/mnras/stx034}

\bibitem[{{Mancera Pi{\~n}a} {et~al.}(2020){Mancera Pi{\~n}a}, {Fraternali},
  {Oman}, {Adams}, {Bacchini}, {Marasco}, {Oosterloo}, {Pezzulli}, {Posti},
  {Leisman}, {Cannon}, {di Teodoro}, {Gault}, {Haynes}, {Reiter}, {Rhode},
  {Salzer}, \& {Smith}}]{mancera_pina2020}
{Mancera Pi{\~n}a}, P.~E., {Fraternali}, F., {Oman}, K.~A., {et~al.} 2020,
  \mnras, 495, 3636, \dodoi{10.1093/mnras/staa1256}

\bibitem[{{Mann} {et~al.}(2015){Mann}, {Feiden}, {Gaidos}, {Boyajian}, \& {von
  Braun}}]{mann2015}
{Mann}, A.~W., {Feiden}, G.~A., {Gaidos}, E., {Boyajian}, T., \& {von Braun},
  K. 2015, \apj, 804, 64, \dodoi{10.1088/0004-637X/804/1/64}

\bibitem[{{Mart{\'\i}n-Navarro}
  {et~al.}(2019{\natexlab{a}}){Mart{\'\i}n-Navarro}, {Romanowsky}, {Brodie},
  {Ferr{\'e}-Mateu}, {Alabi}, {Forbes}, {Sharina}, {Villaume}, {Pandya}, \&
  {Martinez-Delgado}}]{mnavarro2019}
{Mart{\'\i}n-Navarro}, I., {Romanowsky}, A.~J., {Brodie}, J.~P., {et~al.}
  2019{\natexlab{a}}, \mnras, 484, 3425, \dodoi{10.1093/mnras/stz252}

\bibitem[{{Mart{\'\i}n-Navarro}
  {et~al.}(2019{\natexlab{b}}){Mart{\'\i}n-Navarro}, {Romanowsky}, {Brodie},
  {Ferr{\'e}-Mateu}, {Alabi}, {Forbes}, {Sharina}, {Villaume}, {Pandya}, \&
  {Martinez-Delgado}}]{martin-navarro2019}
---. 2019{\natexlab{b}}, \mnras, 484, 3425, \dodoi{10.1093/mnras/stz252}

\bibitem[{{Mercado} {et~al.}(2020){Mercado}, {Bullock}, {Boylan-Kolchin},
  {Moreno}, {Wetzel}, {El-Badry}, {Graus}, {Fitts}, {Hopkins}, \&
  {Faucher-Gigu{\`e}re}}]{mercado2020}
{Mercado}, F.~J., {Bullock}, J.~S., {Boylan-Kolchin}, M., {et~al.} 2020, arXiv
  e-prints, arXiv:2009.01241.
\newblock \doarXiv{2009.01241}

\bibitem[{{Milone} {et~al.}(2011){Milone}, {Sansom}, \&
  {S{\'a}nchez-Bl{\'a}zquez}}]{milone2011}
{Milone}, A. D.~C., {Sansom}, A.~E., \& {S{\'a}nchez-Bl{\'a}zquez}, P. 2011,
  \mnras, 414, 1227, \dodoi{10.1111/j.1365-2966.2011.18457.x}

\bibitem[{{Mistani} {et~al.}(2016){Mistani}, {Sales}, {Pillepich},
  {Sanchez-Janssen}, {Vogelsberger}, {Nelson}, {Rodriguez-Gomez}, {Torrey}, \&
  {Hernquist}}]{mistani2016}
{Mistani}, P.~A., {Sales}, L.~V., {Pillepich}, A., {et~al.} 2016, \mnras, 455,
  2323, \dodoi{10.1093/mnras/stv2435}

\bibitem[{{Morrissey} {et~al.}(2012){Morrissey}, {Matuszewski}, {Martin},
  {Moore}, {Adkins}, {Epps}, {Bartos}, {Cabak}, {Cowley}, {Davis}, {Delacroix},
  {Fucik}, {Hilliard}, {James}, {Kaye}, {Lingner}, {Neill}, {Pistor},
  {Phillips}, {Rockosi}, \& {Weber}}]{morrissey2012}
{Morrissey}, P., {Matuszewski}, M., {Martin}, C., {et~al.} 2012, in Society of
  Photo-Optical Instrumentation Engineers (SPIE) Conference Series, Vol. 8446,
  \procspie, 844613, \dodoi{10.1117/12.924729}

\bibitem[{{Morrissey} {et~al.}(2018){Morrissey}, {Matuszewski}, {Martin},
  {Neill}, {Epps}, {Fucik}, {Weber}, {Darvish}, {Adkins}, {Allen}, {Bartos},
  {Belicki}, {Cabak}, {Callahan}, {Cowley}, {Crabill}, {Deich}, {Delecroix},
  {Doppman}, {Hilyard}, {James}, {Kaye}, {Kokorowski}, {Kwok}, {Lanclos},
  {Milner}, {Moore}, {O'Sullivan}, {Parihar}, {Park}, {Phillips}, {Rizzi},
  {Rockosi}, {Rodriguez}, {Salaun}, {Seaman}, {Sheikh}, {Weiss}, \&
  {Zarzaca}}]{morrissey2018}
{Morrissey}, P., {Matuszewski}, M., {Martin}, D.~C., {et~al.} 2018, \apj, 864,
  93, \dodoi{10.3847/1538-4357/aad597}

\bibitem[{{Mowla} {et~al.}(2017){Mowla}, {van Dokkum}, {Merritt}, {Abraham},
  {Yagi}, \& {Koda}}]{mowla2017}
{Mowla}, L., {van Dokkum}, P., {Merritt}, A., {et~al.} 2017, \apj, 851, 27,
  \dodoi{10.3847/1538-4357/aa961b}

\bibitem[{{M{\"u}ller} {et~al.}(2020){M{\"u}ller}, {Marleau}, {Duc}, {Habas},
  {Fensch}, {Emsellem}, {Poulain}, {Lim}, {Agnello}, {Durrell}, {Paudel},
  {S{\'a}nchez-Janssen}, \& {van der Burg}}]{muller2020}
{M{\"u}ller}, O., {Marleau}, F.~R., {Duc}, P.-A., {et~al.} 2020, \aap, 640,
  A106, \dodoi{10.1051/0004-6361/202038351}

\bibitem[{{Papastergis} {et~al.}(2017){Papastergis}, {Adams}, \&
  {Romanowsky}}]{papastergis2017}
{Papastergis}, E., {Adams}, E.~A.~K., \& {Romanowsky}, A.~J. 2017, \aap, 601,
  L10, \dodoi{10.1051/0004-6361/201730795}

\bibitem[{{Peng} \& {Lim}(2016)}]{peng2016}
{Peng}, E.~W., \& {Lim}, S. 2016, \apjl, 822, L31,
  \dodoi{10.3847/2041-8205/822/2/L31}

\bibitem[{{Peng} {et~al.}(2008){Peng}, {Jord{\'a}n}, {C{\^o}t{\'e}},
  {Takamiya}, {West}, {Blakeslee}, {Chen}, {Ferrarese}, {Mei}, {Tonry}, \&
  {West}}]{peng2008}
{Peng}, E.~W., {Jord{\'a}n}, A., {C{\^o}t{\'e}}, P., {et~al.} 2008, \apj, 681,
  197, \dodoi{10.1086/587951}

\bibitem[{{Peng} {et~al.}(2015){Peng}, {Maiolino}, \& {Cochrane}}]{peng2015}
{Peng}, Y., {Maiolino}, R., \& {Cochrane}, R. 2015, \nat, 521, 192,
  \dodoi{10.1038/nature14439}

\bibitem[{P\'erez \& Granger(2007)}]{PER-GRA:2007}
P\'erez, F., \& Granger, B.~E. 2007, Computing in Science and Engineering, 9,
  21, \dodoi{10.1109/MCSE.2007.53}

\bibitem[{{Ramos-Almendares} {et~al.}(2020){Ramos-Almendares}, {Sales},
  {Abadi}, {Doppel}, {Muriel}, \& {Peng}}]{ra2020}
{Ramos-Almendares}, F., {Sales}, L.~V., {Abadi}, M.~G., {et~al.} 2020, \mnras,
  493, 5357, \dodoi{10.1093/mnras/staa551}

\bibitem[{{Revaz} \& {Jablonka}(2018)}]{rj2018}
{Revaz}, Y., \& {Jablonka}, P. 2018, \aap, 616, A96,
  \dodoi{10.1051/0004-6361/201832669}

\bibitem[{{Rong} {et~al.}(2017){Rong}, {Guo}, {Gao}, {Liao}, {Xie}, {Puzia},
  {Sun}, \& {Pan}}]{rong2017}
{Rong}, Y., {Guo}, Q., {Gao}, L., {et~al.} 2017, \mnras, 470, 4231,
  \dodoi{10.1093/mnras/stx1440}

\bibitem[{{Rong} {et~al.}(2020){Rong}, {Mancera Pi{\~n}a}, {Tempel}, {Puzia},
  \& {De Rijcke}}]{rong2020}
{Rong}, Y., {Mancera Pi{\~n}a}, P.~E., {Tempel}, E., {Puzia}, T.~H., \& {De
  Rijcke}, S. 2020, \mnras, 498, L72, \dodoi{10.1093/mnrasl/slaa129}

\bibitem[{{Ruiz-Lara} {et~al.}(2018){Ruiz-Lara}, {Beasley},
  {Falc{\'o}n-Barroso}, {Rom{\'a}n}, {Pinna}, {Brook}, {Di Cintio},
  {Mart{\'\i}n-Navarro}, {Trujillo}, \& {Vazdekis}}]{rl2018}
{Ruiz-Lara}, T., {Beasley}, M.~A., {Falc{\'o}n-Barroso}, J., {et~al.} 2018,
  \mnras, 478, 2034, \dodoi{10.1093/mnras/sty1112}

\bibitem[{{Saifollahi} {et~al.}(2020){Saifollahi}, {Trujillo}, {Beasley},
  {Peletier}, \& {Knapen}}]{saifollahi2020}
{Saifollahi}, T., {Trujillo}, I., {Beasley}, M.~A., {Peletier}, R.~F., \&
  {Knapen}, J.~H. 2020, arXiv e-prints, arXiv:2006.14630.
\newblock \doarXiv{2006.14630}

\bibitem[{{Sales} {et~al.}(2020){Sales}, {Navarro}, {Pe{\~n}afiel}, {Peng},
  {Lim}, \& {Hernquist}}]{sales2020}
{Sales}, L.~V., {Navarro}, J.~F., {Pe{\~n}afiel}, L., {et~al.} 2020, \mnras,
  494, 1848, \dodoi{10.1093/mnras/staa854}

\bibitem[{Salvatier~J.(2016)}]{pymc3}
Salvatier~J., Wiecki~T.V., F.~C. 2016, PearJ Computer Science

\bibitem[{{S{\'a}nchez-Bl{\'a}zquez} {et~al.}(2006){S{\'a}nchez-Bl{\'a}zquez},
  {Peletier}, {Jim{\'e}nez- Vicente}, {Cardiel}, {Cenarro},
  {Falc{\'o}n-Barroso}, {Gorgas}, {Selam}, \& {Vazdekis}}]{sb2006a}
{S{\'a}nchez-Bl{\'a}zquez}, P., {Peletier}, R.~F., {Jim{\'e}nez- Vicente}, J.,
  {et~al.} 2006, \mnras, 371, 703, \dodoi{10.1111/j.1365-2966.2006.10699.x}

\bibitem[{{Sandage} \& {Binggeli}(1984)}]{sandage1984}
{Sandage}, A., \& {Binggeli}, B. 1984, \aj, 89, 919, \dodoi{10.1086/113588}

\bibitem[{{Schiavon}(2007)}]{schiavon2007}
{Schiavon}, R.~P. 2007, \apjs, 171, 146, \dodoi{10.1086/511753}

\bibitem[{{Schiavon} {et~al.}(2013){Schiavon}, {Caldwell}, {Conroy}, {Graves},
  {Strader}, {MacArthur}, {Courteau}, \& {Harding}}]{schiavon2013}
{Schiavon}, R.~P., {Caldwell}, N., {Conroy}, C., {et~al.} 2013, \apjl, 776, L7,
  \dodoi{10.1088/2041-8205/776/1/L7}

\bibitem[{{Schiavon} {et~al.}(2005){Schiavon}, {Rose}, {Courteau}, \&
  {MacArthur}}]{schiavon2005}
{Schiavon}, R.~P., {Rose}, J.~A., {Courteau}, S., \& {MacArthur}, L.~A. 2005,
  The Astrophysical Journal Supplement Series, 160, 163, \dodoi{10.1086/431148}

\bibitem[{{Serra} \& {Trager}(2007)}]{serra2007}
{Serra}, P., \& {Trager}, S.~C. 2007, \mnras, 374, 769,
  \dodoi{10.1111/j.1365-2966.2006.11188.x}

\bibitem[{{Simon}(2019)}]{simon2019}
{Simon}, J.~D. 2019, \araa, 57, 375,
  \dodoi{10.1146/annurev-astro-091918-104453}

\bibitem[{{Singh} {et~al.}(2019){Singh}, {Zaritsky}, {Donnerstein}, \&
  {Spekkens}}]{singh2019}
{Singh}, P.~R., {Zaritsky}, D., {Donnerstein}, R., \& {Spekkens}, K. 2019, \aj,
  157, 212, \dodoi{10.3847/1538-3881/ab16f2}

\bibitem[{{Sivakoff} {et~al.}(2007){Sivakoff}, {Jord{\'a}n}, {Sarazin},
  {Blakeslee}, {C{\^o}t{\'e}}, {Ferrarese}, {Juett}, {Mei}, \&
  {Peng}}]{sivakoff2007}
{Sivakoff}, G.~R., {Jord{\'a}n}, A., {Sarazin}, C.~L., {et~al.} 2007, \apj,
  660, 1246, \dodoi{10.1086/513094}

\bibitem[{{Smith} {et~al.}(2009){Smith}, {Lucey}, {Hudson}, {Allanson},
  {Bridges}, {Hornschemeier}, {Marzke}, \& {Miller}}]{smith2009}
{Smith}, R.~J., {Lucey}, J.~R., {Hudson}, M.~J., {et~al.} 2009, \mnras, 392,
  1265, \dodoi{10.1111/j.1365-2966.2008.14180.x}

\bibitem[{{Somalwar} {et~al.}(2020){Somalwar}, {Greene}, {Greco}, {Huang},
  {Beaton}, {Goulding}, \& {Lancaster}}]{somalwar2020}
{Somalwar}, J.~J., {Greene}, J.~E., {Greco}, J.~P., {et~al.} 2020, \apj, 902,
  45, \dodoi{10.3847/1538-4357/abb1b2}

\bibitem[{{Thomas} {et~al.}(2011){Thomas}, {Maraston}, \&
  {Johansson}}]{thomas2011}
{Thomas}, D., {Maraston}, C., \& {Johansson}, J. 2011, \mnras, 412, 2183,
  \dodoi{10.1111/j.1365-2966.2010.18049.x}

\bibitem[{{Tolstoy} {et~al.}(2009){Tolstoy}, {Hill}, \& {Tosi}}]{tolstoy2009}
{Tolstoy}, E., {Hill}, V., \& {Tosi}, M. 2009, \araa, 47, 371,
  \dodoi{10.1146/annurev-astro-082708-101650}

\bibitem[{{Tolstoy} {et~al.}(2004){Tolstoy}, {Irwin}, {Helmi}, {Battaglia},
  {Jablonka}, {Hill}, {Venn}, {Shetrone}, {Letarte}, {Cole}, {Primas},
  {Francois}, {Arimoto}, {Sadakane}, {Kaufer}, {Szeifert}, \&
  {Abel}}]{tolstoy2004}
{Tolstoy}, E., {Irwin}, M.~J., {Helmi}, A., {et~al.} 2004, \apjl, 617, L119,
  \dodoi{10.1086/427388}

\bibitem[{{Tremmel} {et~al.}(2020){Tremmel}, {Wright}, {Brooks}, {Munshi},
  {Nagai}, \& {Quinn}}]{tremmel2020}
{Tremmel}, M., {Wright}, A.~C., {Brooks}, A.~M., {et~al.} 2020, \mnras, 497,
  2786, \dodoi{10.1093/mnras/staa2015}

\bibitem[{{Trujillo-Gomez} {et~al.}(2020){Trujillo-Gomez}, {Kruijssen},
  {Keller}, \& {Reina-Campos}}]{trujillogomez2020arXiv}
{Trujillo-Gomez}, S., {Kruijssen}, J.~M.~D., {Keller}, B.~W., \&
  {Reina-Campos}, M. 2020, arXiv e-prints, arXiv:2010.05930.
\newblock \doarXiv{2010.05930}

\bibitem[{{van der Burg} {et~al.}(2016){van der Burg}, {Muzzin}, \&
  {Hoekstra}}]{vdburg2016}
{van der Burg}, R. F.~J., {Muzzin}, A., \& {Hoekstra}, H. 2016, \aap, 590, A20,
  \dodoi{10.1051/0004-6361/201628222}

\bibitem[{Van Der~Walt {et~al.}(2011)Van Der~Walt, Colbert, \&
  Varoquaux}]{van2011numpy}
Van Der~Walt, S., Colbert, S.~C., \& Varoquaux, G. 2011, Computing in Science
  \& Engineering, 13, 22

\bibitem[{{van Dokkum} {et~al.}(2019{\natexlab{a}}){van Dokkum}, {Danieli},
  {Abraham}, {Conroy}, \& {Romanowsky}}]{vd2019b}
{van Dokkum}, P., {Danieli}, S., {Abraham}, R., {Conroy}, C., \& {Romanowsky},
  A.~J. 2019{\natexlab{a}}, \apjl, 874, L5, \dodoi{10.3847/2041-8213/ab0d92}

\bibitem[{{van Dokkum} {et~al.}(2016){van Dokkum}, {Abraham}, {Brodie},
  {Conroy}, {Danieli}, {Merritt}, {Mowla}, {Romanowsky}, \& {Zhang}}]{vd2016}
{van Dokkum}, P., {Abraham}, R., {Brodie}, J., {et~al.} 2016, \apjl, 828, L6,
  \dodoi{10.3847/2041-8205/828/1/L6}

\bibitem[{{van Dokkum} {et~al.}(2017){van Dokkum}, {Abraham}, {Romanowsky},
  {Brodie}, {Conroy}, {Danieli}, {Lokhorst}, {Merritt}, {Mowla}, \&
  {Zhang}}]{vd2017b}
{van Dokkum}, P., {Abraham}, R., {Romanowsky}, A.~J., {et~al.} 2017, \apjl,
  844, L11, \dodoi{10.3847/2041-8213/aa7ca2}

\bibitem[{{van Dokkum} {et~al.}(2018{\natexlab{a}}){van Dokkum}, {Danieli},
  {Cohen}, {Merritt}, {Romanowsky}, {Abraham}, {Brodie}, {Conroy}, {Lokhorst},
  {Mowla}, {O'Sullivan}, \& {Zhang}}]{vd2018}
{van Dokkum}, P., {Danieli}, S., {Cohen}, Y., {et~al.} 2018{\natexlab{a}},
  \nat, 555, 629, \dodoi{10.1038/nature25767}

\bibitem[{{van Dokkum} {et~al.}(2018{\natexlab{b}}){van Dokkum}, {Cohen},
  {Danieli}, {Kruijssen}, {Romanowsky}, {Merritt}, {Abraham}, {Brodie},
  {Conroy}, {Lokhorst}, {Mowla}, {O'Sullivan}, \& {Zhang}}]{vd2018b}
{van Dokkum}, P., {Cohen}, Y., {Danieli}, S., {et~al.} 2018{\natexlab{b}},
  \apjl, 856, L30, \dodoi{10.3847/2041-8213/aab60b}

\bibitem[{{van Dokkum} {et~al.}(2019{\natexlab{b}}){van Dokkum}, {Wasserman},
  {Danieli}, {Abraham}, {Brodie}, {Conroy}, {Forbes}, {Martin}, {Matuszewski},
  {Romanowsky}, \& {Villaume}}]{vd2019}
{van Dokkum}, P., {Wasserman}, A., {Danieli}, S., {et~al.} 2019{\natexlab{b}},
  \apj, 880, 91, \dodoi{10.3847/1538-4357/ab2914}

\bibitem[{{van Dokkum} {et~al.}(2015){van Dokkum}, {Romanowsky}, {Abraham},
  {Brodie}, {Conroy}, {Geha}, {Merritt}, {Villaume}, \& {Zhang}}]{vd2015}
{van Dokkum}, P.~G., {Romanowsky}, A.~J., {Abraham}, R., {et~al.} 2015, \apjl,
  804, L26, \dodoi{10.1088/2041-8205/804/1/L26}

\bibitem[{{van Zee} {et~al.}(2004){van Zee}, {Skillman}, \&
  {Haynes}}]{vanzee2004}
{van Zee}, L., {Skillman}, E.~D., \& {Haynes}, M.~P. 2004, \aj, 128, 121,
  \dodoi{10.1086/421368}

\bibitem[{{VandenBerg} {et~al.}(2013){VandenBerg}, {Brogaard}, {Leaman}, \&
  {Casagrande}}]{vandenberg2013}
{VandenBerg}, D.~A., {Brogaard}, K., {Leaman}, R., \& {Casagrande}, L. 2013,
  \apj, 775, 134, \dodoi{10.1088/0004-637X/775/2/134}

\bibitem[{{Vargas} {et~al.}(2014){Vargas}, {Geha}, \& {Tollerud}}]{vargas2014}
{Vargas}, L.~C., {Geha}, M.~C., \& {Tollerud}, E.~J. 2014, \apj, 790, 73,
  \dodoi{10.1088/0004-637X/790/1/73}

\bibitem[{{Vazdekis} {et~al.}(2015){Vazdekis}, {Coelho}, {Cassisi},
  {Ricciardelli}, {Falc{\'o}n-Barroso}, {S{\'a}nchez-Bl{\'a}zquez}, {La
  Barbera}, {Beasley}, \& {Pietrinferni}}]{vazdekis2015}
{Vazdekis}, A., {Coelho}, P., {Cassisi}, S., {et~al.} 2015, \mnras, 449, 1177,
  \dodoi{10.1093/mnras/stv151}

\bibitem[{{Villaume} {et~al.}(2017){Villaume}, {Conroy}, {Johnson}, {Rayner},
  {Mann}, \& {van Dokkum}}]{villaume2017a}
{Villaume}, A., {Conroy}, C., {Johnson}, B., {et~al.} 2017, \apjs, 230, 23,
  \dodoi{10.3847/1538-4365/aa72ed}

\bibitem[{{Villaume} {et~al.}(2020){Villaume}, {Foreman-Mackey}, {Romanowsky},
  {Brodie}, \& {Strader}}]{villaume2020}
{Villaume}, A., {Foreman-Mackey}, D., {Romanowsky}, A.~J., {Brodie}, J., \&
  {Strader}, J. 2020, \apj, 900, 95, \dodoi{10.3847/1538-4357/aba616}

\bibitem[{{Villaume} {et~al.}(2019){Villaume}, {Romanowsky}, {Brodie}, \&
  {Strader}}]{villaume2019}
{Villaume}, A., {Romanowsky}, A.~J., {Brodie}, J., \& {Strader}, J. 2019, \apj,
  879, 45, \dodoi{10.3847/1538-4357/ab24d7}

\bibitem[{{Virtanen} {et~al.}(2020){Virtanen}, {Gommers}, {Oliphant},
  {Haberland}, {Reddy}, {Cournapeau}, {Burovski}, {Peterson}, {Weckesser},
  {Bright}, {van der Walt}, {Brett}, {Wilson}, {Jarrod Millman}, {Mayorov},
  {Nelson}, {Jones}, {Kern}, {Larson}, {Carey}, {Polat}, {Feng}, {Moore}, {Vand
  erPlas}, {Laxalde}, {Perktold}, {Cimrman}, {Henriksen}, {Quintero}, {Harris},
  {Archibald}, {Ribeiro}, {Pedregosa}, {van Mulbregt}, \&
  {Contributors}}]{2020SciPy-NMeth}
{Virtanen}, P., {Gommers}, R., {Oliphant}, T.~E., {et~al.} 2020, Nature
  Methods, 17, 261, \dodoi{https://doi.org/10.1038/s41592-019-0686-2}

\bibitem[{{Wasserman} {et~al.}(2019){Wasserman}, {van Dokkum}, {Romanowsky},
  {Brodie}, {Danieli}, {Forbes}, {Abraham}, {Martin}, {Matuszewski},
  {Villaume}, {Tamanas}, \& {Profumo}}]{wasserman2019}
{Wasserman}, A., {van Dokkum}, P., {Romanowsky}, A.~J., {et~al.} 2019, arXiv
  e-prints, arXiv:1905.10373.
\newblock \doarXiv{1905.10373}

\bibitem[{{Wheeler} {et~al.}(2019){Wheeler}, {Hopkins}, {Pace},
  {Garrison-Kimmel}, {Boylan-Kolchin}, {Wetzel}, {Bullock}, {Kere{\v{s}}},
  {Faucher-Gigu{\`e}re}, \& {Quataert}}]{wheeler2019}
{Wheeler}, C., {Hopkins}, P.~F., {Pace}, A.~B., {et~al.} 2019, \mnras, 490,
  4447, \dodoi{10.1093/mnras/stz2887}

\bibitem[{{Wright} {et~al.}(2020){Wright}, {Tremmel}, {Brooks}, {Munshi},
  {Nagai}, {Sharma}, \& {Quinn}}]{wright2020}
{Wright}, A.~C., {Tremmel}, M., {Brooks}, A.~M., {et~al.} 2020, arXiv e-prints,
  arXiv:2005.07634.
\newblock \doarXiv{2005.07634}

\bibitem[{{Yagi} {et~al.}(2016){Yagi}, {Koda}, {Komiyama}, \&
  {Yamanoi}}]{yagi2016}
{Yagi}, M., {Koda}, J., {Komiyama}, Y., \& {Yamanoi}, H. 2016, \apjs, 225, 11,
  \dodoi{10.3847/0067-0049/225/1/11}

\bibitem[{{Yozin} \& {Bekki}(2015)}]{yozin2015}
{Yozin}, C., \& {Bekki}, K. 2015, \mnras, 452, 937,
  \dodoi{10.1093/mnras/stv1073}

\end{thebibliography}
\bibliographystyle{aasjournal}

\appendix





\section{Estimates of Stellar Masses}\label{ap:masses}

\subsection{This work}

DF44's luminosity comes from the $g$ band total integrated magnitude from \citet{vd2015} ($-15.7^{+0.2}_{-0.2}$). We estimated the stellar mass for DF44 using the $M/L$ ratio inferred from our stellar population synthesis modeling of the integrated spectrum ($M/L_r = 1.21$) for a stellar mass ${\rm log}_{10} M_* = 8.48$. The K-correction made a negligible difference to the estimated stellar mass.

\subsection{The Coma dwarfs}

The Coma dwarfs included in Figure~\ref{fig:gp_feh} are in part from \citet{smith2009}, which details how the luminosities were computed. We multiplied the luminosities with $M/L$ estimated from their reported ages and metallicities using the Flexible Stellar Population Synthesis \citep[FSPS, python-FSPS;][]{conroy2010, pythonfsps} models. 
For the others we used the stellar masses published in \citet{gu2018a}, \citet{rl2018}, and \citet{ferre-mateu2018}.

\subsection{The Coma UDGs}

The Coma UDGs included in Figure \ref{fig:gp_feh} include those in \citet{ferre-mateu2018}, \citet{rl2018}, \citet{gu2018a}, and \citet{chilingarian2019}. For both we use the estimated stellar masses provided in those papers. For the UDGs included in \citet{chilingarian2019} we used the $M/L$ and absolute $r$-band magnitudes those authors provided to estimate stellar mass.

\subsection{Local Group dwarfs}

All dwarfs represented by filled gray circles in Figure \ref{fig:gp_feh} come from \citet{kirby2013}. 
The stellar mass for Andromeda XIX was estimated from the published luminosity included in \citet{collins2020} and a $M/L$ estimated from the published metallicity and assuming an age of $10$ Gyr.

\end{document}